\documentclass{siamart171218}

\usepackage{lipsum}
\usepackage{amsfonts}
\usepackage{graphicx}
\usepackage{epstopdf}
\usepackage{algorithmic}
\usepackage{graphicx, amsmath, amssymb, mathtools, dirtytalk, bm, float} 
\usepackage[symbol]{footmisc}
\usepackage{color}
\usepackage{hyperref}
\usepackage{comment}
\usepackage{xr}
\usepackage{cite}

\newcommand{\RR}{{\mathbb R}}
\newcommand{\ZZ}{{\mathbb Z}}

\def\Xint#1{\mathchoice
   {\XXint\displaystyle\textstyle{#1}}%
   {\XXint\textstyle\scriptstyle{#1}}%
   {\XXint\scriptstyle\scriptscriptstyle{#1}}%
   {\XXint\scriptscriptstyle\scriptscriptstyle{#1}}%
   \!\int}
\def\XXint#1#2#3{{\setbox0=\hbox{$#1{#2#3}{\int}$}
     \vcenter{\hbox{$#2#3$}}\kern-.5\wd0}}

\def\dashint{\Xint-}

\ifpdf
  \DeclareGraphicsExtensions{.eps,.pdf,.png,.jpg}
\else
  \DeclareGraphicsExtensions{.eps}
\fi

\title{A Christov integrating factor method for approximating rational PDE solutions}
\author{Corresponding author: Troy I. Johnson \thanks{Department of Mathematics, University of Colorado Colorado Springs, Colorado Springs, United States of America
  (\email{tjohnso3@uccs.edu}})
\and Justin T. Cole\thanks{Department of Mathematics, University of Colorado Colorado Springs, Colorado Springs, United States of America
  (\email{jcole13@uccs.edu}.)}}

\usepackage{amsopn}

\begin{document}

\maketitle

\begin{abstract}
Rational solutions of partial differential equations (PDEs) are notoriously difficult to approximate via spectral Fourier methods due to their algebraically slow decay rate.
In this work we discuss approximating rational PDE solutions in a basis of orthogonal functions known as the Christov functions, allowing for the computation of its spectrum via the fast Fourier transform. An explicit fourth-order integrating factor method is derived. The effectiveness of the method is tested on several nonlinear PDEs. Relative to a straightforward  pseudo-spectral approximation, the integrating factor method supports larger regions of stability resulting in a more efficient scheme.
\end{abstract}



\section{Introduction}
\label{intro_sec}

    The standard way of numerically approximating localized \\functions on the real line is by Fourier methods. Even though the function is not strictly periodic, typically one can take a sufficiently large computational window so that computational problem {\it looks} periodic. For sufficiently smooth solutions, this approach yields exponentially decaying Fourier modes which can be rapidly computed via  fast Fourier transform \cite{Fornberg1996,Canuto2006}. The key assumption here is that the function reaches its boundary conditions within the computational domain. Exponentially localized solutions, like certain soliton solutions
in nonlinear PDEs \cite{Yang2010}, are ideal candidates for this procedure. 

On the other hand, localized rational functions struggle  under this approach. The issue is a practical one since any square integrable rational function with poles off the real axis has a Fourier transform that decays exponentially in wavenumber. The issue is that numerically, one must choose a {\it finite} computational interval to work on. If the function is truncated too early, the resulting periodic extension contains 
small kinks in the tails. As a result, the corresponding Fourier modes decay algebraically slow, severely limiting the numerical accuracy. 

This is not a new problem and over the years researchers have devised several methods for dealing with this issue.
One popular approach is that of coordinate mapping \cite{Grosch1977}. Here one maps  the  real line to a finite interval by some appropriate coordinate transformation. Then, one could use another basis, like Chebyshev, to approximate a function and its derivatives \cite{Boyd1987}. The price one must pay is that this introduces variable coefficients to the approximation. 
Another idea is to use the Hermite function basis \cite{Boyd1980}. Intuitively, this  approach makes sense because the Hermite functions form an orthogonal basis on the real line. However,  the truncated series approximation only converges spectrally fast if the function decays exponentially fast at $\pm \infty$ \cite{Canuto2006}. 

An alternative approach we look to continue developing in this work is the application of the Christov (Ch) functions \cite{Christov1982,Boyd1990,Iserles2020} as an approximation basis. The Ch functions are a family of rational functions that are orthogonal on the real line. Furthermore, via a coordinate transformation, it is possible to rapidly compute the modal coefficients via a modified fast Fourier transform. The main limitation of this basis is that it clusters grid points near the origin. As a result, for localized functions who are {\it not} concentrated
near the origin, their corresponding Ch coefficients decay still exponential, but not as rapidly.

In this article we develop a versatile integrating-factor (IF) 
Christov approach for approximating rational PDE solutions.  A thorough derivation of the  integrating-factor method and its properties can be found in  \cite{Boyd2001}.  For rational functions that are square-integrable, we develop a method that is spectrally accurate in space and fourth-order accurate in time; we integrate the IF equations with fourth-order Runge-Kutta method. The exponential convergence rate is related to the width of an analytic strip in the complex plane of a Fourier-like problem \cite{Paley1934}. For benchmark  simulations considered in this work, we are consistently able to achieve seven to eight  decimal places of accuracy in only a few seconds on a standard personal laptop. In a previous work, we introduced a split-step Christov method \cite{Cole2025}. The advantage of the IF method is that it is easier to accommodate generic nonlinearities relative to the split-step method. An advantage that both the split-step and integrating-factor methods have over a direct pseudo-spectral method is a significantly larger region of stability in the selection of timestep.

To highlight the versatility of this method, we approximate rational solutions of the  nonlinear Schr\"odinger (NLS), modified Korteweg-de Vries (mKdV), and \\Benjamin-Ono (BO) equations. These equations are relevant in the field of nonlinear waves and support soliton-like  rational solutions. One notable example is the Peregrine soliton solution of the NLS equation (which we approximate below), a well-known rogue wave model for deep water waves \cite{Kharif2003,Chabchoub2011} and fiber optics \cite{Agrawal2013}.

The outline of the work is as follows. In Section 2 we introduce the Ch functions, the Ch expansion and some of their properties. In Section 3 we review the rapid decay of the coefficients of the Ch expansion. In Section 4, spectral Galerkin differentiation and Hilbert transform matrices are constructed to approximate derivatives and the Hilbert transform, respectively. In Section 5 we develop a novel spectral integrating-factor method for approximating rational solutions of nonlinear PDEs. We test the effectiveness of this method on several physically important
PDEs in Section 6. We conclude in Section 7.  Several appendices with extra technical notes, figures and results are also included.

\section{The Christov (Ch) functions}
\label{Ch_sec}

\begin{figure}
 \centering
 \includegraphics[scale=0.3]{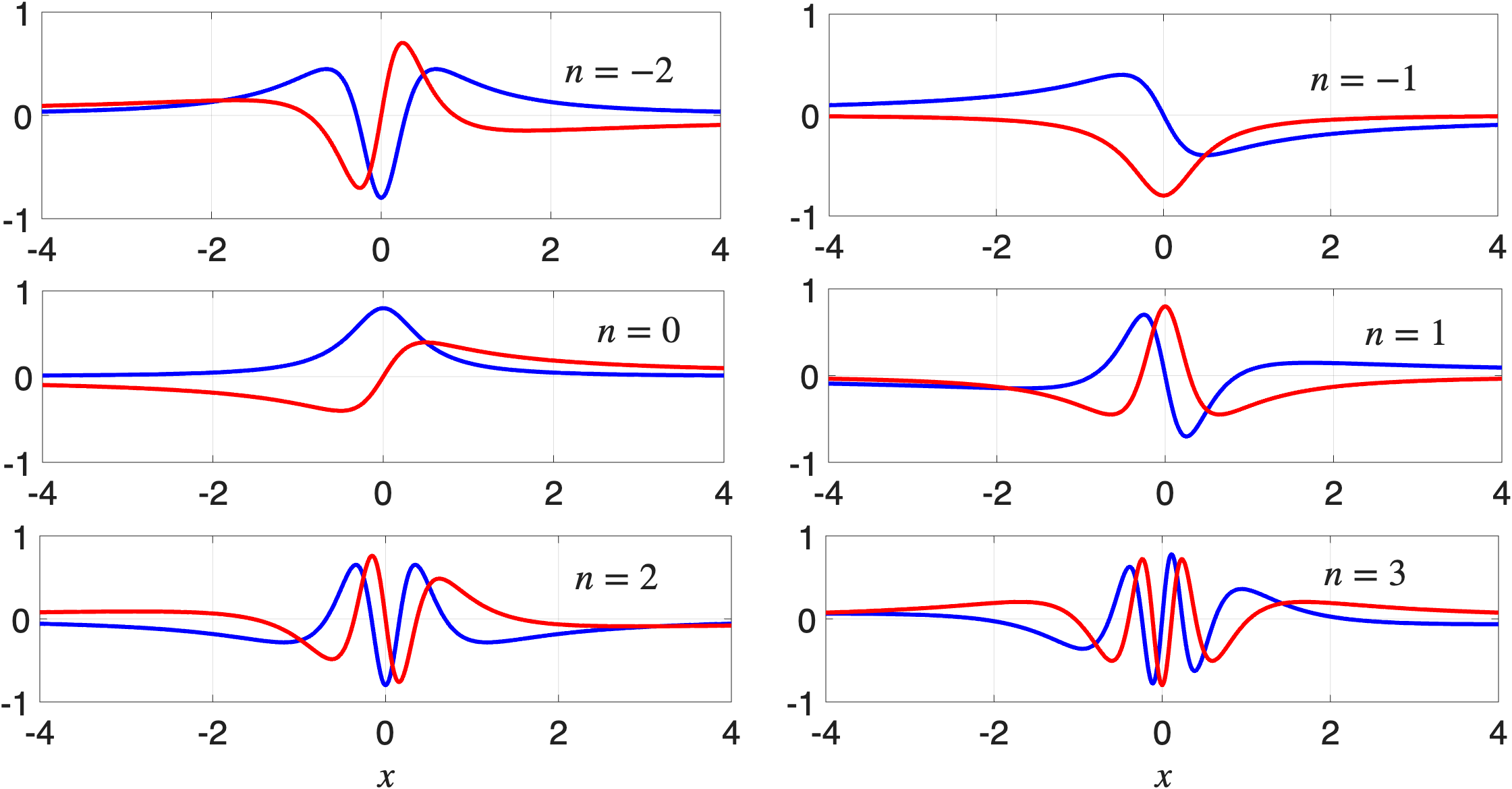}
 \caption{The real (blue curves) and imaginary (red curves)  parts of $\phi_n(x)$ defined in Eq.~(\ref{Ch_fcn_define}) for $n \in \{-2, \dots, 3 \}$. } 
 \label{plot_Ch_fcns}
 \end{figure}

The Christov functions were first presented in 1982 by their namesake to approximate solutions of a nonlinear  boundary value problems \cite{Christov1982}.
Define the family of 
Christov rational functions 
 \begin{equation}
 \label{Ch_fcn_define}
   \phi_n(x)=\sqrt{\frac{2}{\pi}}i^n\left(\frac{1+2ix}{1-2ix}\right)^n\frac{1}{1-2ix} , ~~~~~ n \in \ZZ.
 \end{equation} 
  Observe that $\phi_n(x)$ is well-defined for $x \in \mathbb{R}$.
  Below is a list of some properties of these functions; a few of the functions are plotted in Fig.~\ref{plot_Ch_fcns}. 

 \begin{itemize}

\item {\it Poles}
The $n^{\rm th}$ Ch function $\phi_n(x)$ has a degree $n+1$ pole at $x = -i/2$ when $n$ is positive,   and  a degree $n$ pole at $x = i/2$ and a simple pole at $x = -i/2$ when $n $ is nonpositive.

 \item  {\it Oscillations} 
  The Ch modes increase in frequency near the origin as $|n| \rightarrow \infty$. One can see this by noting the rational function that lies on the unit circle, $\frac{1 + 2 i x}{1 - 2 i x} $, can be expressed as $e^{ i  \theta} $. Then  $\phi_n(x)$ contains a factor of the form $e^{ i n \theta} = \cos (n \theta) + i \sin (n \theta)$ whose  frequency increases with $|n|$. 
 
  \item {\it Orthogonality} 
  Define the complex $L^2(\mathbb{R})$ inner product
\begin{equation}
\label{define_inner_prod}
\langle f,g \rangle =  \int_{-\infty}^{\infty}f^*(x)g(x)dx ,
\end{equation}
where $^*$ denotes  complex conjugation. The Ch functions are mutually 
orthonormal \cite{Iserles2020}, that is 
\begin{equation}
\label{ortho_define}
\langle \phi_m,\phi_n \rangle = \int_{-\infty}^{\infty}\phi^*_m(x)\phi_n(x)dx= \delta_{mn} ,
\end{equation}
where  $ \delta_{mn}$ denotes the Kronecker delta function. The coefficients  in (\ref{Ch_fcn_define}) are chosen to also ensure unit norm, i.e. $|| \phi_n || = \sqrt{\langle \phi_n, \phi_n \rangle} = 1$, for each $n$. Furthermore, the functions form a complete  basis in $L^2(\RR)$ \cite{Higgins1977}.

\item  {\it Derivative recurrence relation} 
The Ch functions satisfy the 
skew-symmetric recurrence relation \cite{Christov1982,Iserles2020} 
\begin{equation}
\label{deriv_recurrence}
\frac{d}{dx}\phi_n(x)=-n\phi_{n-1}(x)+i(2n+1)\phi_n(x)+(n+1)\phi_{n+1}(x).
\end{equation}
This formula naturally leads to tridiagonal spectral differentiation matrices. Higher-order derivatives can be obtained recursively by differentiating this formula.

\item  {\it Hilbert transform eigenfunction} 
The Ch functions are eigenfunctions of the Hilbert transform.
The Hilbert transform is defined by 
\begin{equation}
\label{Hil_tran}
    \mathcal{H}[u](x)=\frac{1}{\pi} \dashint_{-\infty}^{\infty}\frac{u(y)}{y-x}dy,
\end{equation}
where the integral is defined in the Cauchy principal values sense
 \begin{equation*}
 \dashint_{-\infty}^{\infty}f(x)dx=\underset{\epsilon \to 0^+}{\lim} \left[\int_{-\infty}^{b-\epsilon}f(x)dx \hspace{0.05in}+ \hspace{0.05in} \int_{b+\epsilon}^{\infty}f(x)dx \right] .
\end{equation*}
The Ch functions satisfy the eigenvalue problem \cite{Weideman1995}
\begin{equation}
\label{Ch_hilbert}
\mathcal{H}[\phi_n](x) =  i\hspace{0.05in}\text{sgn}(n)\phi_n(x) ~~{\rm where}~~ \text{sgn}(n)= \begin{cases} 
      1 & \text{if} \hspace{0.3in} n \geq 0 \\
      -1 & \text{if} \hspace{0.3in} n<0 \\
   \end{cases} .
\end{equation}

 \end{itemize}

\noindent
{\bf Transformation to Fourier series}\\

Throughout this work we consider expressing a function in terms of the  Ch functions, that is
\begin{equation}
\label{Ch_expansion}
 f(x) = \sum_{n = - \infty}^{\infty} \Check{f}_n\phi_n(x),\hspace{0.1in} \text{where} \hspace{0.1in} \Check{f}_n= \langle \phi_n , f  \rangle = \int_{-\infty}^{\infty} \phi^*_n(x) f(x)  ~ dx ,
 \end{equation}
for the Ch functions given in Eq.~(\ref{Ch_fcn_define}). This expansion is unique and the coefficients are obtained by projecting onto an arbitrary mode and exploiting the orthogonality (\ref{ortho_define}). Utilizing a natural change of variable, this Ch expansion can be related to the Fourier series of a periodic function. This is useful in numerical simulations since it  allows one to utilize the Fast Fourier transform (FFT) to compute the Ch coefficients in Eq.~(\ref{Ch_expansion}). 

To begin, notice that the term $\frac{1+2ix}{1-2ix}$ lies on the complex unit circle. This motivates the the coordinate transformation 
\begin{equation}
\label{x_theta_transform}
e^{i\theta}=\frac{1+2ix}{1-2ix} ~~~  \Leftrightarrow ~~~ x=\frac{1}{2}\tan\left( \frac{\theta}{2} \right)  ,
\end{equation}
which defines a map between the real line $x \in (- \infty, \infty)$ and finite interval $\theta \in (- \pi , \pi)$.
Applying this change of variable to the Ch functions in Eq.~(\ref{Ch_fcn_define}) yields
\begin{equation}
 \label{Ch_theta} 
 \phi_n\left( \frac{1}{2}\tan\left( \frac{\theta}{2} \right) \right)=\sqrt{\frac{2}{\pi}}i^ne^{i n \theta}\frac{1}{1-i\tan \left(\frac{\theta}{2} \right)} = \sqrt{\frac{2}{\pi}}i^n e^{i\theta \left(n+\frac{1}{2} \right)}\cos\left(\frac{\theta}{2}\right).
 \end{equation}

  Observe that the Ch function in Eq.~(\ref{Ch_theta}) is $2\pi$-periodic in  $\theta$. Utilizing the transformation in Eq.~(\ref{x_theta_transform}), the projection integral in Eq.~(\ref{Ch_expansion}) becomes
 \begin{equation} 
\label{Ch_coeff_theta}
  \Check{f}_n=\int_{-\infty}^{\infty} \phi^*_n(x)  f(x) dx =\frac{(-i)^n}{2\sqrt{2\pi}}\int_{-\pi}^{\pi} f\left(\frac{1}{2}\tan\left(\frac{\theta}{2} \right)\right)\left(1-i\tan \left(\frac{\theta}{2} \right) \right) e^{-i n \theta } d\theta .
  \end{equation}
 Thus, there is a direct relationship between the Ch coefficients and the Fourier coefficients of the (periodic) function  $ f\left(\frac{1}{2}\tan\left(\frac{\theta}{2} \right)\right)\left(1-i\tan \left(\frac{\theta}{2} \right) \right) $ on the interval $(- \pi , \pi )$. 
 
The Ch coefficients corresponding to the function $f(x)$ are computed via

\begin{equation}  
\label{Ch_Fourier_abbrev}
 \check{\textbf{f}} =\mathcal{MF}[f(x)]= \textbf{k}\sqrt{\frac{\pi}{2}}\mathcal{F}\left[ f \left(\frac{1}{2}\tan \left(\frac{\theta}{2} \right) \right)\left(1-i\tan \left(\frac{\theta}{2} \right) \right) \right] ,
  \end{equation}
where $\check{\textbf{f}} = \left( \dots , \check{f}_{-1}, \check{f}_0, \check{f}_1 , \dots \right)^T$ and $\textbf{k} = \text{diag}\left( \dots , (-i)^{-1}, (-i)^{0}, (-i)^{1} , \dots \right),$    
  and the function is summed (see Eqs.~(\ref{Ch_expansion}) and (\ref{Ch_theta})) by 
 \begin{equation} 
 \label{MIFFT}
f(x) = \mathcal{MF}^{-1}[\Check{f}_n]=\sqrt{\frac{2}{\pi}}\frac{\mathcal{F}^{-1}\left[ i^n\Check{f}_n  \right]}{1-i\tan \left(\frac{\theta}{2} \right)} ,
  \end{equation} 
 where $\mathcal{F}$ and $\mathcal{F}^{-1}$  denote the Fourier and inverse Fourier (semi-discrete) transforms, respectively. 
 We refer to  $\mathcal{MF}$ and $\mathcal{MF}^{-1}$ as the  {\it modified Fourier} and {\it inverse modified Fourier transforms}, respectively, which we 
 utilize to compute the Ch coefficients of a given function.  To be clear, the Ch coefficients are computed directly
 through Fourier transforms in the $\theta$ variable. Note that the divisor, $1-i\tan \left(\frac{\theta}{2}\right)$, in Eq.~(\ref{MIFFT}), is included because Eq.~(\ref{Ch_Fourier_abbrev}) is computing the  Fourier transform of the function $ f\left(\frac{1}{2}\tan\left(\frac{\theta}{2} \right)\right)\left(1-i\tan \left(\frac{\theta}{2} \right) \right) =f\left(x\right)\left(1-2ix\right)$, so one recovers the function $f(x)$ by dividing out the term, $1-i\tan \left(\frac{\theta}{2}\right)$. 
 The discrete versions of the Fourier transform are naturally computed using the fast Fourier transform (FFT) algorithm 
 (see \cite{Cole2025}).\\  

\noindent
{\it Parseval relation} \\\\
The series expansion in Eq.~(\ref{Ch_expansion}) exhibits an equality of the spatial and spectral $L^2$ norms. This fact follows immediately from the orthogonality (\ref{ortho_define}), 
\begin{equation}
\label{parseval}
\int_{-\infty}^{\infty} |f(x)|^2 dx  
=  \sum_{n, n'}  \check{f}_n^* \check{f}_{n'}   \int_{-\infty}^{\infty} \phi^*_n(x) \phi_{n'}(x) dx = \sum_{n = - \infty}^{\infty} \left|\check{f}_n \right|^2 .
\end{equation}

\section{Decay rate of Ch coefficients}
\label{decay_sect}

In this work we focus on rational functions that are square integrable and analytic on the real line. In particular, we focus on rational functions of the form $\frac{p(x)}{q(x)}$ where $p(x)$ and $q(x)$ are polynomials of degree $r$ and $s$, respectively, with $r < s$ and any common factors of zero canceled out. We may also consider the case $r = s$ when $\frac{p(x)}{q(x)}$ approaches a nonzero boundary value of $\frac{p_{\infty}}{q_{ \infty}}  $ as $x \rightarrow \pm \infty$. This latter class of rational of functions is clearly not in $L^2(\mathbb{R})$, however the shifted function $\frac{p(x)}{q(x)} - \frac{p_{\infty}}{q_{\infty}}  $ does decay to zero at infinity and  is in $L^2(\mathbb{R})$. By incorporating this shift, we are still able to apply a Ch  approximation like Eq.~(\ref{Ch_expansion}); see Sec.~\ref{num_pde_sec}.

An attractive feature of the Ch basis is the exponential decay of its coefficients  for analytic and bounded rational functions \cite{Weideman1995}. Consider  the transformed Ch coefficients in Eq.~(\ref{Ch_coeff_theta}), where the integral is in terms of the angular variable $\theta$. Consider complexifying $\theta$ within some analytic strip
  $\left|{\rm Re} ~\theta \right| \le  \pi $ and $ \left| {\rm Im}~\theta\right| <  \kappa$, such that the function $f\left( \frac{1}{2} \tan \left( \frac{\theta}{2} \right) \right) \left( 1 -  i \tan \left( \frac{\theta}{2} \right) \right) $ is uniformly bounded within the strip. Then the Ch coefficients decay like 
  $|\Check{f}_n| \approx O( e^{- |n \kappa|})$, that is, exponentially fast \cite{Paley1934}. More generally, as long as this function is analytic, then the Ch coefficients should decay exponentially fast in $n$ \cite{Weideman1995,Iserles2020}. 

Suppose a function $f(z)$ has a pole at $z=\xi+i\eta$.  Then the transformation in Eq.~(\ref{x_theta_transform}) implies a pole at $\theta=2\tan^{-1}(2(\xi+i\eta))$. We then note that the complex form of the inverse tangent is given by, $\tan^{-1}(z)=-\frac{i}{2}\log\left(\frac{1+iz}{1-iz}\right)$, and the complex logarithm is, $\log(z)=\log|z|+i \arg(z)$, and so the distance from this pole in $\theta$ to the real axis is given by

\begin{equation}
    \label{decay_rate}
   \kappa =  \text{Im}\left[2\tan^{-1}(2( \xi +i\eta))\right]= \begin{cases}\frac{1}{2}\log\left(\frac{(1+2\eta)^2+4\xi^2}{(1-2\eta)^2+4\xi^2}\right) & \text{if } \eta>0\\
   -\frac{1}{2}\log\left(\frac{(1-2\eta)^2+4\xi^2}{(1+2\eta)^2+4\xi^2}\right) & \text{if } \eta<0      
   \end{cases}
\end{equation}
Recall the property of logarithms;  $\log\left(\frac{1}{a}\right)=-\log(a)$ for $a>0 $. Hence, the value of $\kappa$ in Eq.~(\ref{decay_rate}) that corresponds to the pole of $f(z)$ closest to the real axis fixes the asymptotic convergence rate,
$|\Check{f}_n| \sim e^{- |n \kappa|}$. We note a special case, if $\xi=0$ and $\eta=\pm\frac{1}{2}$ then the poles of $f$ match the poles of the Christov functions. As an example, for the rational function $(1+4x^2)^{-1}$ 
the sum given in (\ref{Ch_expansion}) is finite and $\frac{1}{1+4x^2}=\frac{1}{2(1+2ix)}+\frac{1}{2(1-2ix)}=\Check{f}_{-1}\phi_{-1}(x)+\Check{f}_0\phi_0(x)$, where $\Check{f}_{-1}=\frac{i\sqrt{\pi}}{2\sqrt{2}}$ and $\Check{f}_{0}=\frac{\sqrt{\pi}}{2\sqrt{2}}$. 

Consider the illuminating example 
\begin{equation}
    \label{decay_funk}
    f(x)=\frac{1}{1+(x-a)^2}, ~~~~a\in\RR,
\end{equation}
which has poles at $z =  a \pm i$ when we complexify $x$. This is a single-humped positive function with a peak at $x = a$. We examine this function in particular since it corresponds to a translated or shifted rational function. It's profile matches a few of the solutions we examine in Sec.~\ref{num_pde_sec}, including the mKdV and BO equations.  
For any $a \in \RR$, the distance between the poles to the real axis in $\theta$ in Eq.~(\ref{decay_rate}) is given by
\begin{equation}
    \label{ex_decay_rate}
   \kappa =  \text{Im}\left[2\tan^{-1}(2a\pm 2i)\right]=\frac{1}{2}\log\left(\frac{(1\pm 2)^2+4a^2}{(1\mp 2)^2+4a^2}\right) = \pm \frac{1}{2}\log\left(\frac{9+4a^2}{1+4a^2}\right) .
\end{equation}

 \begin{figure}
    \centering
   \includegraphics[scale=0.325]{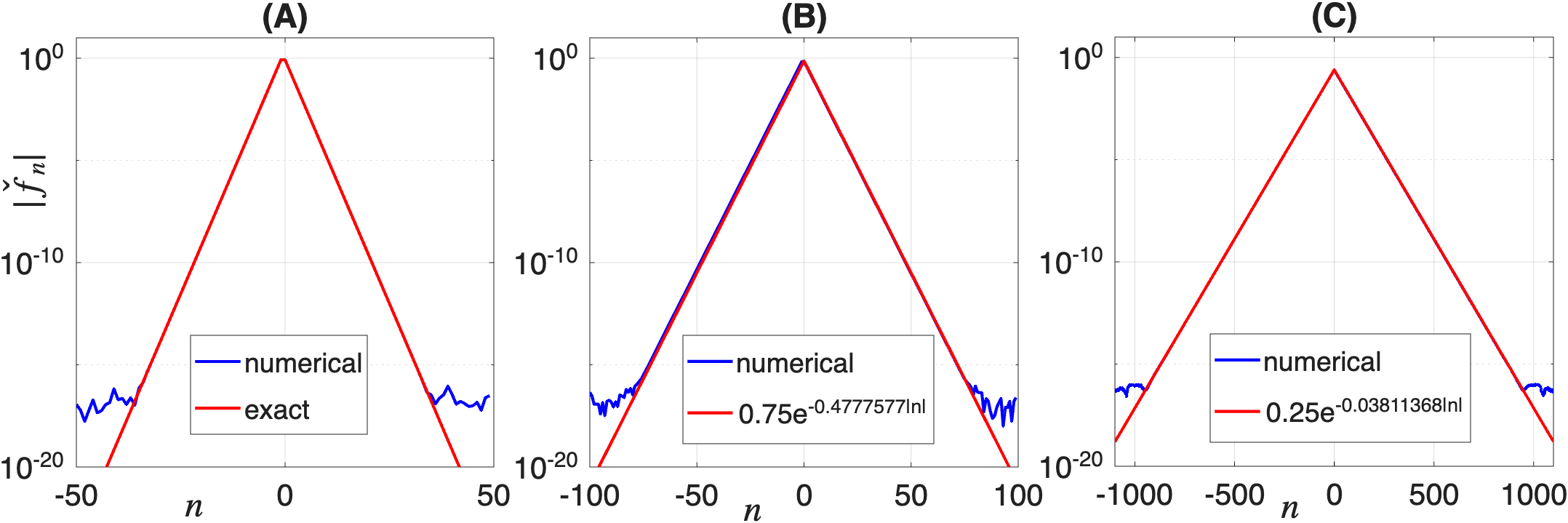}
   \caption{(A) Numerically computed Ch coefficients for the function $f(x) = (1+x^2)^{-1}$ in blue, the exact Ch coefficients  (\ref{rational_expand_exact}) are in red. (B) Numerically computed Ch coefficients for the function $f(x) = (1+(x-1)^2)^{-1}$ in blue, asymptotic decay rate of Ch coefficients  in red. (C) Numerically computed Ch coefficients for the function $f(x) = (1+(x+5)^2)^{-1}$ in blue, asymptotic decay rate of Ch coefficients  in red. Notice that 
   the Ch coefficients decay exponentially fast.\label{Ch_coeff_decay_numerical}}
   \end{figure}

Take $a=0$, so that $f(z) = (1 + z^2 )^{-1}$. Under  transformation (\ref{x_theta_transform}), the poles of the function are located at $z = \pm i$.  Since the distance from the real line to the pole is  $\log(3) \approx 1.0986$, the Fourier-Ch coefficients defined in Eq.~(\ref{Ch_coeff_theta}) are predicted from Eq.~(\ref{ex_decay_rate}) to decay exponentially with the form $|\check{f}_n| \sim   e^{- \log(3)|n| }=3^{-|n|}$ as $n \rightarrow \pm \infty$, which agrees with the exact
formula for this series
\begin{equation}
\label{rational_expand_exact}
\frac{1}{1 + x^2} = -  \sqrt{2 \pi} \sum_{n = - \infty}^{-1} \frac{i^n}{3^{-n}} \phi_{n}(x) +  \sqrt{2 \pi} \sum_{n = 0}^{\infty} \frac{i^n}{3^{n+1}} \phi_n(x) ;
\end{equation}
see Fig.~\ref{Ch_coeff_decay_numerical}(A).

We now consider shifting the function (\ref{decay_funk}) away from the origin by taking  $a=1$. Observe that $f(x) = (1 + (x-1)^2)^{-1}$ has poles at $z  =1\pm i$ in the original  coordinates. In the angular coordinates, the distance from the real line to the imaginary part of the poles is obtained from Eq.~(\ref{ex_decay_rate}) and found to be $\kappa = \frac{1}{2}\log\left(\frac{13}{5}\right)\approx0.4778$. As a result, 
the Ch-Fourier coefficients decay like $|\check{f}_n| \sim  e^{-0.4778 |n| }$ as $n \rightarrow \pm \infty$. Notice that the decay rate decreases as the peak of the function (\ref{decay_funk}) at $x = a$ moves away from the origin. This is further highlighted if we take $a=-5$ in (\ref{decay_funk}) which has poles at $x  = 5 \pm i$. The distance from the real angular line to the imaginary part of the poles is  $\kappa = \frac{1}{2}\log\left(\frac{109}{101}\right)\approx0.0381$. Hence, the Ch-Fourier coefficients decay like $|\check{f}_n| \sim  e^{-0.0381 |n| }$ as $n \rightarrow \pm \infty$.

 A comparison between the asymptotic rates discussed above and the numerically computed values is shown in Fig.~\ref{Ch_coeff_decay_numerical}; there is good agreement.
We note that as $|a| \to \infty$ the distance in Eq.~(\ref{ex_decay_rate}) approaches $\kappa = \frac{1}{2}\log\left(\frac{9+4a^2}{1+4a^2}\right) \rightarrow 0$, so as the magnitude of $a$ increases we expect slower decay of the  Ch coefficients. For the magnitude of Ch coefficients at $10^{-k}$ for $k \in \ZZ^+$ we expect $|n| \approx 2k/\log_{10}\left(\frac{9+4a^2}{1+4a^2}\right)$. We observe that for $a=0,1$ the Ch coefficients of each function decay exponentially fast, reaching round-off error at around $N = 68$ and $N=156$ respectively; a modest amount of grid points. But for $a=-5$ we reach round off error at $N=1934$; significantly more grid points. More examples of coefficient decay are given in Appendix \ref{decay_app}.

What these  examples highlight is that the Ch coefficients decay exponentially fast for analytic rational functions, which indicates they are good candidates for a spectral method since spectral accuracy is directly related to the decay rate of the coefficients (see next section). In turn, this section shows that the position of the function affects the decay rate. Intuitively, it makes sense that the approximation works best when the solution is located near the origin, where the spatial grid points (in $x$) are clustered. These insights inform some of our conclusions later when we analyze its performance in solving PDEs with traveling wave solutions.

\section{Spectral differentiation and the Hilbert transform}
\label{diff_mat_sec}

Let us review the  matrices for differentiation and Hilbert transform. To begin, consider  a function $f(x)$ expressed in terms of Ch functions as in (\ref{Ch_expansion}) whose derivative is 
 \begin{equation}
 \label{series_deriv_recur}
f'(x)   = \sum_{n = - \infty}^{\infty} \Check{f}_n \phi'_n(x)  = \sum_{n = - \infty}^{\infty}  \Check{f}^{(1)}_n \phi_n(x) ,
 \end{equation}
where $\Check{f}^{(1)}_n$ denotes the derivative function coefficients. The second series motivates our desire to express the derivative function in terms of an expansion of Ch functions.
Recall, that the derivative of a Ch function is given by 
the recurrence relation in Eq.~(\ref{deriv_recurrence}).
Substituting this expansion in first series in Eq.~(\ref{series_deriv_recur}) and then shifting the indices results in the following relationship between the coefficients of the derivative and original function
\begin{equation}
\label{deriv_coeff_relate}
\Check{f}^{(1)}_n =  n \Check{f}_{n-1}  + i(2n + 1) \Check{f}_n - (n +1) \Check{f}_{n+1}   .
\end{equation}
A Ch derivative coefficient involves itself $\{n\}$ and nearest two \{$n-1, n+1$\} modes. Alternatively, the derivative coefficients can be computed via matrix multiplication
\begin{equation}
\label{Ch_diff_mat_relate}
\check{\textbf{f}}^{(1)} = \left( \dots , \check{f}^{(1)}_{-1}, \check{f}^{(1)}_0, \check{f}^{(1)}_1 , \dots \right)^T, ~~~~ \check{\textbf{f}}^{(1)} =\mathbb{D}_1 \check{\textbf{f}} , ~~~~ \check{\textbf{f}} = \left( \dots , \check{f}_{-1}, \check{f}_0, \check{f}_1 , \dots \right)^T ,
\end{equation} 
where $\mathbb{D}_1$ is the banded  matrix
\begin{equation*}
\mathbb{D}_1 =   \left[
\begin{array}{*{10}c}
 \ddots&  \ddots&   &  & &   & & & & \\
 \ddots &  \ddots&  \ddots &  & &   & & & & \\
   &  -3 &  -5i  & 2 & &   & & & & \\
   &   &  -2  & -3i & 1 &   & & & & \\
   &   &    & -1 & -i & 0  & & & & \\
   &   &    &  & 0 & i  & -1 & & & \\
    &   &    &  &  & 1  & 3i & -2 & & \\
    &   &    &  &  &   & 2 & 5i &-3 & \\
     &   &    &  &  &   &  & \ddots & \ddots &\ddots \\
     &   &    &  &  &   &  & & \ddots &\ddots 
\end{array} 
\right] .
\end{equation*}
This matrix is tridiagonal (banded) and skew-Hermitian ($\mathbb{D}_1^\dag = - \mathbb{D}_1$), where $^\dag$ denotes the conjugate transpose. 
While it is not as sparse as a (diagonal) Fourier differentiation matrix, its bandwith of 1 means that it is still quite sparse. By differentiating the recurrence relation in Eq.~(\ref{deriv_recurrence}), one can construct higher-order differentiation matrices. The second and third differentiation matrices are given in Appendix (\ref{third_Ch_matrix}). 

Next, we express the Hilbert transform in terms of 
Ch functions, that is
 \begin{equation*}
\mathcal{H}[f][x] =\sum_{n = - \infty}^{\infty}  \Check{f}_n  \mathcal{H}[\phi_n][x]  =\sum_{n = - \infty}^{\infty}  \Check{f}^{(h)}_n \phi_n(x) ,
 \end{equation*}
where $\Check{f}^{(h)}_n$ reflects the application of the Hilbert transform. 
Recall the Ch functions are eigenfunctions of the Hilbert transform, see Eq.~(\ref{Ch_hilbert}). 
Hence, substituting the eigenvalue relation $\mathcal{H}[\phi_n][x] = i \text{sgn}(n) \phi_n(x)$ gives 
\begin{equation}
\label{Ch_hil_coeef}
\Check{f}^{(h)}_n=  i\hspace{0.05in}\text{sgn}(n)\Check{f}_n  , ~~{\rm where}~~ \text{sgn}(n)= \begin{cases} 
      1 & \text{if} \hspace{0.3in} n \geq 0 \\
      -1 & \text{if} \hspace{0.3in} n<0 \\
   \end{cases}   .
\end{equation}

Applying the Hilbert transform to an Ch expansion amounts to self-modal scalar multiplication by by a factor $\pm i$.
Alternatively, the application of the Hilbert transform 
can be computed via matrix multiplication
\begin{equation}
\label{Ch_hil_mat_relate}
\check{\textbf{f}}^{(h)} = \left( \dots , \check{f}^{(h)}_{-1}, \check{f}^{(h)}_0, \check{f}^{(h)}_1 , \dots \right)^T, ~~~~ \check{\textbf{f}}^{(h)} =\mathbb{H} \check{\textbf{f}} , ~~~~ \check{\textbf{f}} = \left( \dots , \check{f}_{-1}, \check{f}_0, \check{f}_1 , \dots \right)^T ,
\end{equation} 
where $\mathbb{H}$ is the diagonal matrix
$\mathbb{H}=  \text{diag}\left[\text{sgn}(n)i\right] $. 
In practice, one would not store the full matrix for $\mathbb{H}$.

 To approximate derivatives or the Hilbert transform via Ch expansions in practice,  
the series in Eq.~(\ref{Ch_expansion}) are truncated to  
$N$ modes: $-N/2, -N/2 + 1 , \dots, N/2-1$, where $N$ is a positive even integer. The Ch coefficients $\Check{f}_n$ of the function $f(x)$ are computed using the modified Fourier transform in Eq.~(\ref{Ch_Fourier_abbrev}). The corresponding differentiation matrices, e.g. Eq.~(\ref{Ch_diff_mat_relate}), or Hilbert transform, i.e. Eq.~(\ref{Ch_hil_mat_relate}), are truncated to an $N \times N$ system and applied. The inverse modified transform in Eq.~(\ref{MIFFT}) is used to recover the approximation in physical space.

\begin{figure}
\centering
   \includegraphics[scale=0.34]{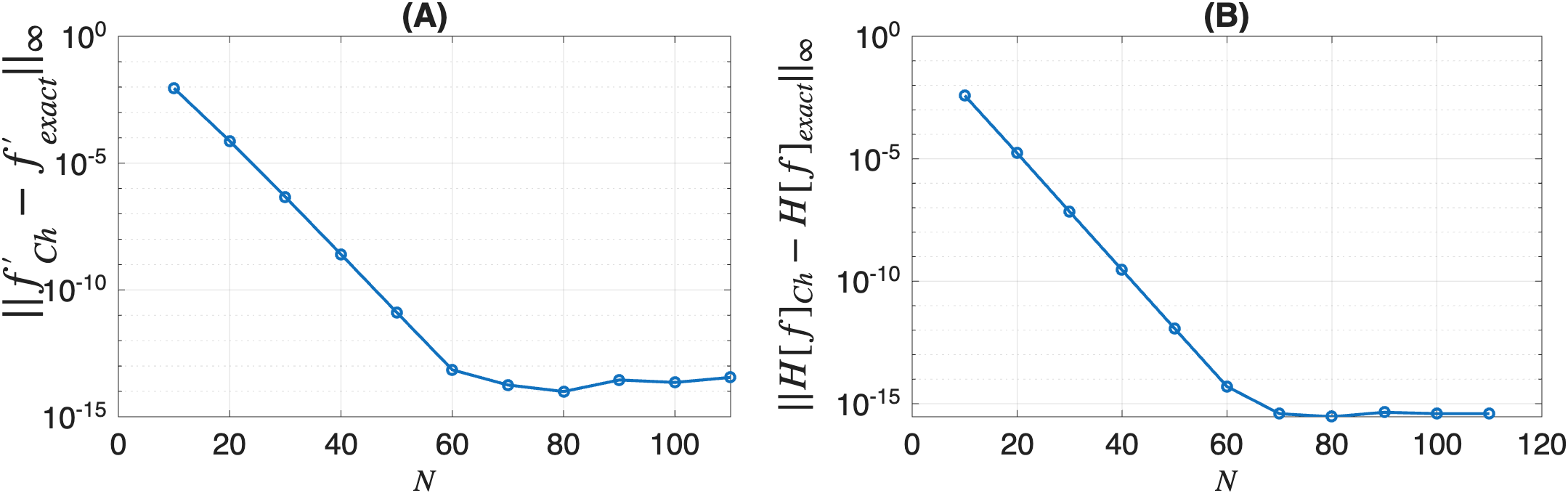}
 \caption{Infinity norm error of the (A) first  Ch derivative and (B) Ch Hilbert transform approximations applied to the rational function $f(x) = (1 + x^2)^{-1}$ as a function of the number of Ch modes, $N$. The error approaches zero exponentially fast. 
 \label{dcon}}
   \end{figure}

A typical set of convergence results is highlighted in Fig.~\ref{dcon}. For the rational function $f(x) = 1/(1 +x^2)$, the Ch approximation of the derivative and Hilbert transform is remarkably effective. The error converges exponentially fast to zero until we reach round-off error effects at around $N = 70$ modes. 

\section{Numerical approximation of time-dependent PDEs}
\label{num_pde_sec}

In this section we  develop an integrating-factor method for approximating rational solutions of time-dependent PDEs. A novel time-stepping approach is implemented using the differentiation and Hilbert transform matrices described in the previous section. An explicit fourth-order Runge-Kutta integrator is used to advance the time step. The effectiveness of the method is highlighted in the next section for several physically important nonlinear PDEs.

We consider the general  nonlinear initial  value problem
\begin{equation}
\label{general_nonlin_PDE}
u_t=\mathcal{L}u+ \mathcal{N}[u] ,  ~~~~~ u_0 (x) = u(x,t_0) , ~~~~~ (x,t) \in \mathbb{R} \times [t_0, T] , 
\end{equation}
where the linear and nonlinear operators are denoted by $\mathcal{L}$ and $\mathcal{N}$, respectively. 
The linear operator is assumed to be a 
differential operator 
of 
the form 
$$ \mathcal{L}= \left(  c_1\frac{\partial}{\partial x} + c_2 \frac{\partial ^2}{\partial x^2}+ c_3 \frac{\partial ^2}{\partial x^3} + \cdots\right) + \mathcal{H} \left(   d_1\frac{\partial}{\partial x} + d_2 \frac{\partial ^2}{\partial x^2}+ d_3 \frac{\partial ^2}{\partial x^3} + \cdots\right)  ~ ,$$
 with   scalar coefficients $  c_1 ,c_2, \dots $ and $ d_1 ,d_2, \dots$, and  Hilbert transform operator $\mathcal{H}$. The nonlinear operator is assumed to depend on some combination of $u$ and its derivatives or Hilbert transform. The boundary conditions (BCs) are assumed to approach some constant value at the infinite boundaries, i.e. $u(x,t) \rightarrow U$ as $x \rightarrow \pm \infty$. For rational functions, this implies $r \le s$, where $r$ ($s$) is the degree of the numerator (denominator) polynomial. Furthermore, we assume nonlinearities with constant and bounded BCs, $N[u] \rightarrow N[U] < \infty$.
 
 To approximate the linear operator in Eq.~(\ref{general_nonlin_PDE}) we use the spectral matrices developed in Section~\ref{diff_mat_sec}. For solutions with zero BCs ($U = 0$), the appropriate truncated differentiation matrix $\mathbb{D}_i$ or Hilbert transform matrix $\mathbb{H}$ is applied to the Ch coefficients calculated in Eq.~(\ref{Ch_Fourier_abbrev}). However, if $U \neq 0$, then $u(x,t) \not\in L^2(\mathbb{R})$ and a direct expansion in terms of the Ch functions is not possible.

To get around this, we take advantage of the basic calculus property $\frac{d u}{d x} = \frac{d}{d x} (u - U ) .$ Both sides of the equation have the same derivative, but the right-hand side function $u - U$ has zero BCs and can be directly expanded in a basis of Ch functions. As a result, we introduce the shifted function $w(x,t) = u(x,t) - U$ to Eq.~(\ref{general_nonlin_PDE}),
\begin{equation}
\label{shift_nonlin_PDE}
w_t=\mathcal{L}w+ \mathcal{N}[w + U], 
\end{equation}
where $w \rightarrow 0$ as $x \rightarrow \pm \infty$.

Then to approximate the nonlinear operator in Eq.~(\ref{shift_nonlin_PDE}) we 
utilize a psuedospectral approach. Terms like 
derivatives or the Hilbert transform are 
approximated by the appropriate operator in spectral space while all other nonlinearities, like $u^p$, are evaluated in physical space. In order to ensure appropriate BCs we 
subtract off 
the nonlinear  BC, similar to the process above, and then apply the modified Fourier transform (\ref{Ch_Fourier_abbrev}), which yields   
\begin{equation}
\label{Ch_nonlin_PDE}
\Check{\bf w}_t=\mathbb{L}\Check{\bf w}+ \mathcal{MF}\Big[\mathcal{N}[w +U]-\mathcal{N}[U]\Big] , 
\end{equation}
where $\Check{\bf w}=\mathcal{MF}[w]$ and 
$$\mathbb{L}=\left( c_1\mathbb{D}_1+c_2\mathbb{D}_2+c_3\mathbb{D}_3+ \cdots  \right) + \mathbb{H} \left( d_1\mathbb{D}_1+d_2\mathbb{D}_2+d_3\mathbb{D}_3+ \cdots  \right) , $$
 with $\mathbb{D}_i$ and $\mathbb{H}$ being the Christov differentiation and Hilbert transform matrices defined in Section \ref{diff_mat_sec} and \ref{third_Ch_matrix}. Clearly, $\mathcal{N}[w + U ]-\mathcal{N}[U] \rightarrow 0$ as $x \rightarrow \pm \infty$. 

Taking the integrating-factor (IF) transformation $\Check{\bf w}=e^{\mathbb{L}t}\Check{\bf v}$ removes the $\mathbb{L}\Check{\bf w}$ term in Eq.~(\ref{Ch_nonlin_PDE}) and yields the system of differential equations


\begin{equation}
\label{general_if_form}
\Check{\bf v}_t  = e^{-\mathbb{L}t}\mathcal{MF}\Big[\mathcal{N}[w + U ]-\mathcal{N}[U]\Big] ,
\end{equation}
 where $w=\mathcal{MF}^{-1}\Big[e^{\mathbb{L}t}\Check{\bf v}\Big]$ and $u=w+U$.  Recall that the nonlinear operator, $\mathcal{N}$, is assumed to depend on some 
product of $u$ and its derivatives or Hilbert transform.  If it depends on the latter, we apply the appropriate operator in spectral space while all nonlinearities, like $u^p$, are evaluated in physical space. As such, we break the nonlinear operator into two parts, $\mathbb{N}_R[u]$, to be applied in real space and $\mathbb{N}_{\text{Ch}}$, which is a matrix that'll be applied in spectral space (see Appendix \ref{IVPs}).

Note that $e^{\mathbb{L}t}$ denotes the matrix exponential evaluation of the spectral linear matrix $\mathbb{L}t$. For a dispersive system, $\mathbb{L}$ is a skew-Hermitian matrix  with purely imaginary spectrum and $e^{\mathbb{L}t}$ is a bounded matrix function. The traditional benefit of IF methods is their enhanced stability properties for explicit integrators 
\cite{Yang2010,Boyd2001}. We numerically integrate by a
fourth-order Runge-Kutta method. In practice, we find the integrating-factor method has a much larger region of stability leading to
run-times that are an order of magnitude less than direct pseudo-spectral Runge-Kutta methods. For example, in Fig.~\ref{nls_converge}(B) the total CPU time ranged from 1-8.5 seconds, whereas using similar parameters in a pseudo-spectral 4th order Runge-Kutta method took over 2 minutes. Pseudo-code is given in Appendix \ref{IVPs}.

Below we test the performance of this method on several nonlinear dispersive PDEs with well-known rational solutions. When we do not have a known solution we 
monitor  the accuracy of the method by conservation laws (see below) and localization of the Ch coefficients. The latter assumption was motivated in Section~\ref{decay_sect} and is important for truncating to a finite number of modes. In practice, we require the magnitude of the first and the last Ch coefficients (the most oscillatory)
to be less than $10^{-3}$. Experience shows that above this, obvious errors will manifest in the numerical approximation.

\section{Examples}
\label{example_Sec}

In this section we apply the Ch-IF method detailed in the previous section to several nonlinear dispersive PDEs with rational solutions. We considered a split-step method in \cite{Cole2025} for the NLS equation. We solve this equation again here with the IF to: (a) highlight the versatility of the IF method and (b) examine the differences between the two methods, e.g. stability and accuracy.

\subsection{Nonlinear Schr\"odinger (NLS) equation}
\label{Ch4_sec}

\begin{figure}
    \centering
   \includegraphics[scale=0.3]{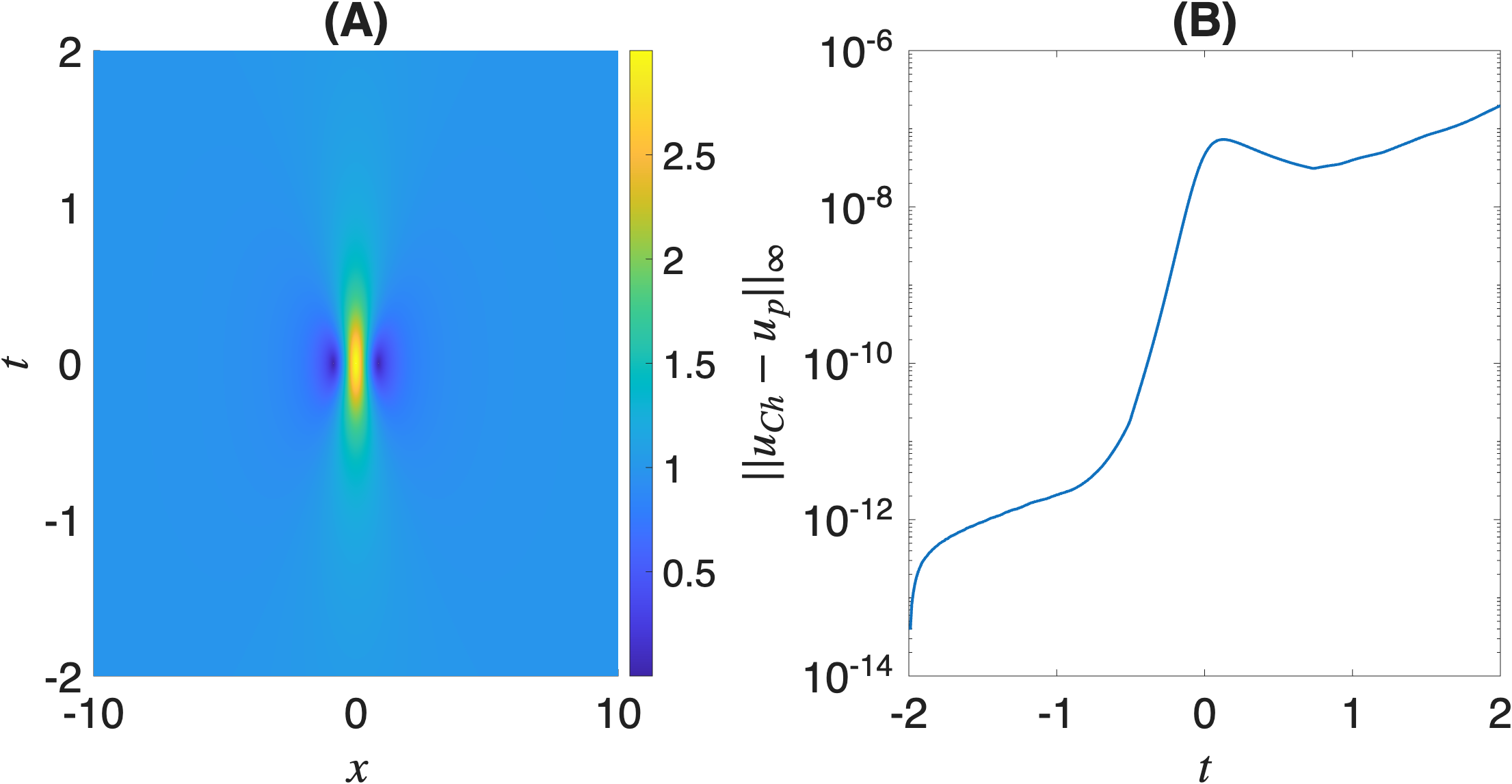}
   \caption{(A) Evolution of $|u(x,t)|$ and (B) corresponding error of the Ch-IF  method for the 
   Peregrine soliton in Eq.~(\ref{peregrine}) in the NLS equation (\ref{nls_eq}) for  $t_0=-2$. 
   The computational parameters are: $-2\leq t \leq2$,
   $N = 300$ Ch modes and time-step $\Delta t=0.001$. The large error growth  can be attributed to the unstable nature of the Peregrine soliton. \label{nls_graph} } 
   \end{figure}

The NLS equation is a physically important and  well-studied integrable equation \cite{Ablowitz2011}   and a good candidate for gauging the performance of this numerical scheme.
Consider the non-dimensionalized nonlinear Schr\"odinger  equation 
\begin{equation}
    \label{nls_eq}
    iu_t+u_{xx}+2(|u|^2-1)u=0 ,  ~~~~~ u_0 (x) = u(x,t_0) , ~~~~ (x,t) \in \mathbb{R} \times [t_0,T] , 
\end{equation}
with boundary conditions $u \to U = 1$ as $x \to \pm \infty$. The extra factor $- 2 u$ is included for convenience; it can be removed by the transformation $u(x,t) = \psi(x,t) e^{-2 i t }$. The NLS equation is used to describe weakly nonlinear quasi-monochormatic wave trains. The well-known Peregrine soliton solution 
\begin{equation}
    \label{peregrine}
    u_p(x,t)=\frac{4x^2-16it+16t^2-3}{4x^2+16t^2+1} , 
\end{equation}
is used to model rogue waves in a variety of physical systems \cite{Chabchoub2011,Calini2019,Ablowitz2021}. The Peregrine soliton is special because it is localized in space and time; see Fig.~\ref{nls_graph}(A).

To numerically solve Eq.~(\ref{nls_eq}) we note that the nonlinear part is in terms of $u$ 
and thus the algorithm given in \ref{IVPs} with 
$$\mathbb{L}=i\mathbb{D}_2,~~\mathbb{N}_R[u]=2i(|u|^2-1)u,~~\mathbb{N}_{\text{Ch}}=\mathbb{I},~~\text{and}~~ U=1, $$

where $\mathbb{D}_2$ is the second differentiation matrix defined in Appendix \ref{third_Ch_matrix} and $\mathbb{I}$ is the identity matrix.

\subsubsection{A convergence study}

\begin{figure}
    \centering
   \includegraphics[scale=0.325]{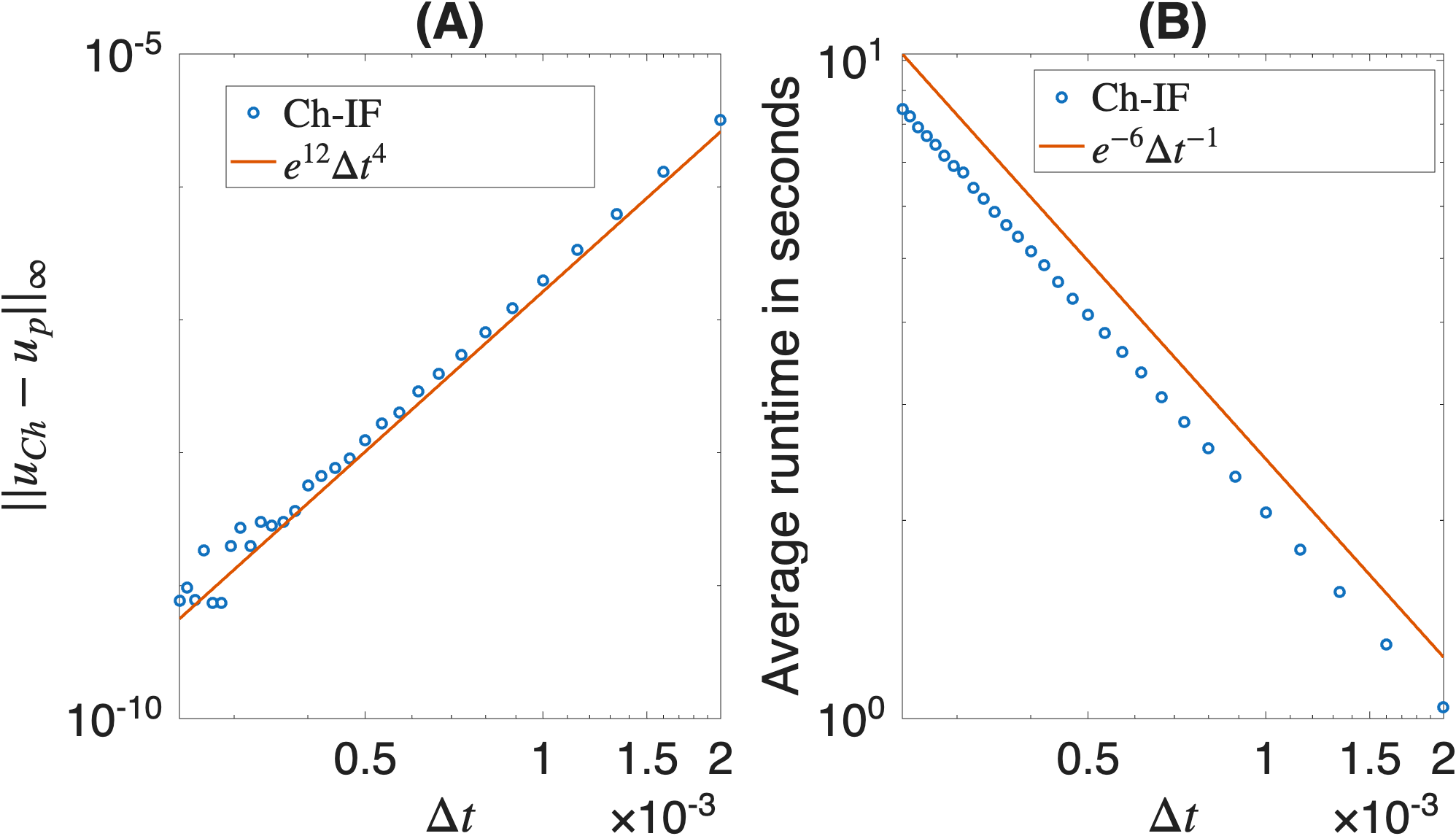}
   \caption{(A) Convergence of 
   Ch-IF approximation for the Peregrine soliton (\ref{peregrine}) solved on the interval $-2 \le t \le 2$. The global error over the space-time window
   for $N = 300$  is shown. The Ch-IF method converges  fourth-order as  $\Delta t $ decreases.   (B) Average CPU runtime as a function of $\Delta t$ and fixed $N=300$. The runtime  increases linearly with $\Delta t^{-1}$.  
   \label{nls_converge}} 
   \end{figure}

Typical results obtained by  solving the NLS equation (\ref{nls_eq})  for the Peregrine soliton solution (\ref{peregrine}) with the Ch-IF method are discussed next. The evolution of a typical Peregrine soliton obtained using this method is shown in Fig.~\ref{nls_graph}(A) over the time interval $[-2,2]$. This interval includes the rogue wave peak  of $|u(0,0)| = 3$ at the space-time origin.


 The global spatial error as a function of time is highlighted in Fig.~\ref{nls_graph}(B). For a time-step size of $\Delta t = 10^{-3}$, the difference between the exact and numerical approximation is $\mathcal{O}(10^{-6})$ or better. In an earlier work we developed a split-step Ch method for integrating NLS-type equations \cite{Cole2025}. For similar parameters, we compare the error in Appendix \ref{splitstep};  we observe the IF method accuracy is comparable.

Next, the  convergence rate and runtime as a function of $\Delta t$ are
discussed.  In Fig.~\ref{nls_converge}(A) the convergence rate of the Ch-IF method applied to  the Peregrine solution (\ref{peregrine}) is shown. A a clear fourth-order convergence rate is observed until other errors become significant.
On the other hand, 
corresponding runtimes are shown in Fig.~\ref{nls_converge}(B). This time encapsulates only  the IF solver portion of the code (no pre or post computations). We observe a roughly linear dependence for this range of computational parameters. For all runs considered in Fig.~\ref{nls_converge}, the total CPU time ranged from 1-8.5 seconds. All runs in this work were performed on a MacBook Pro laptop with  M2 chip and 16GB of memory.  The data values are the average of over 10 runs to mitigate various CPU-related fluctuations in the runtime. In practice we utilize the sparse form of the matrices in order to reduce runtime.

\begin{figure}
    \centering
   \includegraphics[scale=0.35]{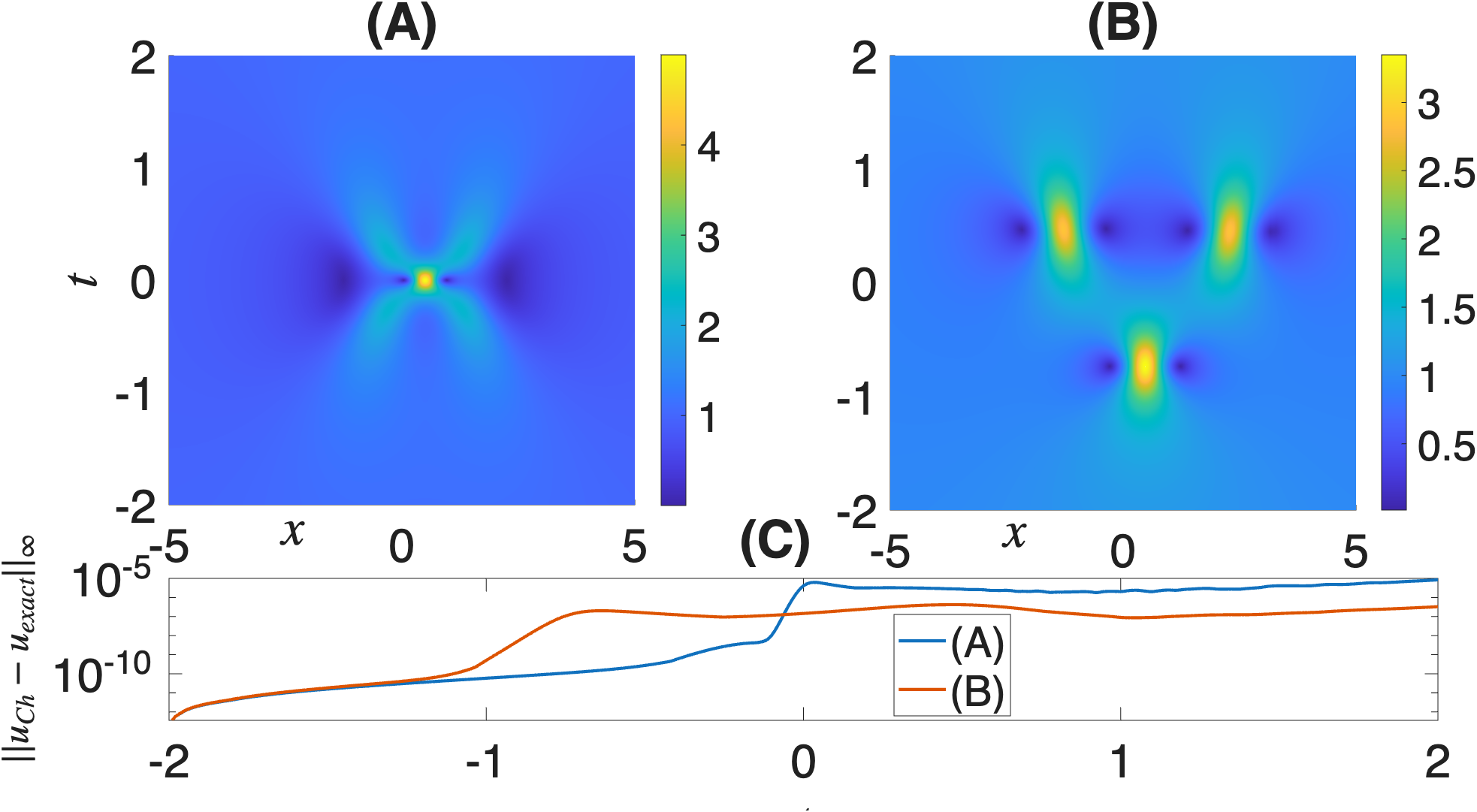}
   \caption{(A) Evolution of  $|u(x,t)|$ for the second-order Peregrine soliton in Eq.~(\ref{peregrine2}) with $a=\frac{-1}{12}$, (B) Evolution of  $|u(x,t)|$ for the second-order Peregrine soliton in Eq.~(\ref{peregrine2}) with $a=\frac{5i}{2}$  and (C) error of of the Ch-IF numerical approximations in the NLS equation (\ref{nls_eq}) for  $t_0=-2$. Computational parameters are: $-2\leq t \leq2$,
   $N = 500$ Ch modes and time-step $\Delta t=0.001$.   \label{nls2_graph} } 
   \end{figure}

To further exhibit the effectiveness of this method, we approximate a higher order Peregrine soliton. 
The second-order Peregrine soliton is given by \cite{Yang2012}

\begin{align}
    \label{peregrine2}
    & u_{p_2}(x,t)=1+\frac{v(x,t)}{w(x,t)} , \\ \nonumber
 & v(x,t)= 24\Big[\big(3x-6x^2+4x^3-2x^4-48t^2
+48xt^2-48x^2t^2-160t^4\big)
\\ \nonumber
 & -it\big(-12+12x-16x^3+8x^4+32t^2-64xt^2+64x^2t^2+128t^4\big)
 \\ \nonumber
 & +6a\big(1-2x+x^2+4it-4ixt-4t^2\big)+6a^*\big(-x^2-4ixt+4t^2\big)\Big],
    \\ \nonumber
      & w(x,t)=\big(9-36x+72x^2-72x^3+72x^4-48x^5+16x^6\big)\\ \nonumber & +96t^2\big(3+3x-4x^3+2x^4)+384t^4\big(5-2x+2x^2\big)+1024t^6  
      \\ \nonumber & +24\big(a+a^*\big)\big(3x^2-2x^3-12t^2+24xt^2\big)
      \\ \nonumber & +48i\big(a-a^*\big)\big(-3t-6xt+6x^2t-8t^3\big)+144|a|^2,
\end{align}

where $a$ is a free complex parameter. The evolution of a second-order soliton obtained using the Ch-IF method is shown in Fig.~\ref{nls2_graph}(A). With $a=- \frac{1}{12}$ the soliton peaks at the origin $(x,t) = (0,0)$, however: (a) there are additional ``x'' arms before and after the peak, and (b) the peak amplitude is 5, larger than that of the first-order soliton in Eq.~(\ref{peregrine}). Fig.~\ref{nls2_graph}(B) shows  evolution of a typical second-order soliton obtained using the Ch-IF  with the complex parameter $a=\frac{5i}{3}$. Here the structure that forms around $t = -1$ bears 
a resemblance to that of the Peregrine soliton (see Fig.~\ref{nls_graph}). Namely, a localized peak of amplitude nearly 3 appears and then suddenly disappears. Afterward, a two-hump mode emerges and then disappears, similar to a modulation instability ``wake'' pattern \cite{Biondini2016,Yuen1978}. Global spatial error as a function of time is highlighted in Fig.~\ref{nls2_graph}(C). 
The difference between the exact and numerical approximations are $\mathcal{O}(10^{-5})$ or better. To obtain these results, the number of Ch modes is increased relative to the fundamental soliton shown in Fig.~\ref{nls_graph}. Like the previous case, we expect this solution to be unstable; this explains the large growth of the error.


\subsubsection{Conserved quantities}
\label{conserved_quant_sec}

For  localized solutions of the NLS equation (\ref{nls_eq}) that decay to zero, the following integrals
\begin{equation}
\label{conserved_quant}
P = \int_{-\infty}^{\infty} |u|^2~ dx , ~~~~~~ M = {\rm Im} \int_{-\infty}^{\infty} u^* u_x ~dx , ~~~~~~ H = \int_{-\infty}^{\infty} \left(|u_x|^2 - |u|^4 \right) dx ,
\end{equation}
denoting power, momentum, and Hamiltonian, respectively, are constants of motion. This numerical method should accurately approximate these physical quantities in addition to the solution itself. 
Note that for the Peregrine solitons in Eqs.~(\ref{peregrine}) and (\ref{peregrine2}), these integrals do not converge.

To approximate the integrals in Eq.~(\ref{conserved_quant}) the integrands are  constructed  as  the product of two Ch expansions, e.g. Eq~(\ref{Ch_expansion}). The reason is that the Ch functions are  in $L^2(\mathbb{R})$, but not $L^1(\mathbb{R})$. So for a generic integral $\int_{-\infty}^{\infty} f^*(x) g(x)  ~dx$, approximation by Ch expansions gives
$$
\left| \int_{-\infty}^{\infty} f^*(x) g(x)  ~dx \right| = \left|\sum_{m} \sum_n \Check{f}^*_m \Check{g}_n \int_{-\infty}^{\infty} \phi^*_m \phi_n ~dx\right| = \left|\sum_{n}  \Check{f}^*_n \Check{g}_n \right| . 
$$
Application of the  Cauchy-Schwarz inequality and Parseval relation (\ref{parseval}) shows that this is bounded quantity, provided $f(x), g(x) \in L^2(\mathbb{R})$.

The  Parseval relation in Eq.~(\ref{parseval}) is used to approximate the power integral $P$ in Eq.~(\ref{conserved_quant}). The derivative term $u_x$  in the momentum, $M$, is evaluated through the spectral differentiation matrix in Eq.~(\ref{Ch_diff_mat_relate}), and so
\begin{equation}
\label{discrete_momentum}
M  = {\rm Im} ~ \check{\bf u}^\dag (\mathbb{D}_1  \check{\bf u}) , ~~~~~ \check{\bf u} = (\cdots, \check{u}_{-1},  \check{u}_{0}, \check{u}_{1}, \cdots )^T .
\end{equation}
Denoting $\rho(x,t) = u^2(x,t)$, the Hamiltonian is evaluated by 
\begin{equation}
\label{discrete_Hamil}
H  =
( \mathbb{D}_1  \Check{\bf u})^\dag ( \mathbb{D}_1  \Check{\bf u})  -  \Check{\bm \rho}^\dag \Check{\bm \rho} , ~~~~~ \Check{\bm \rho} = (\cdots, \check{\rho}_{-1},  \check{\rho}_{0}, \check{\rho}_{1}, \cdots )^T .
\end{equation}
In practice, we truncate these quantities to $N$ modes, where $N \gg 1 $. 

\begin{figure}
    \centering
   \includegraphics[scale=0.325]{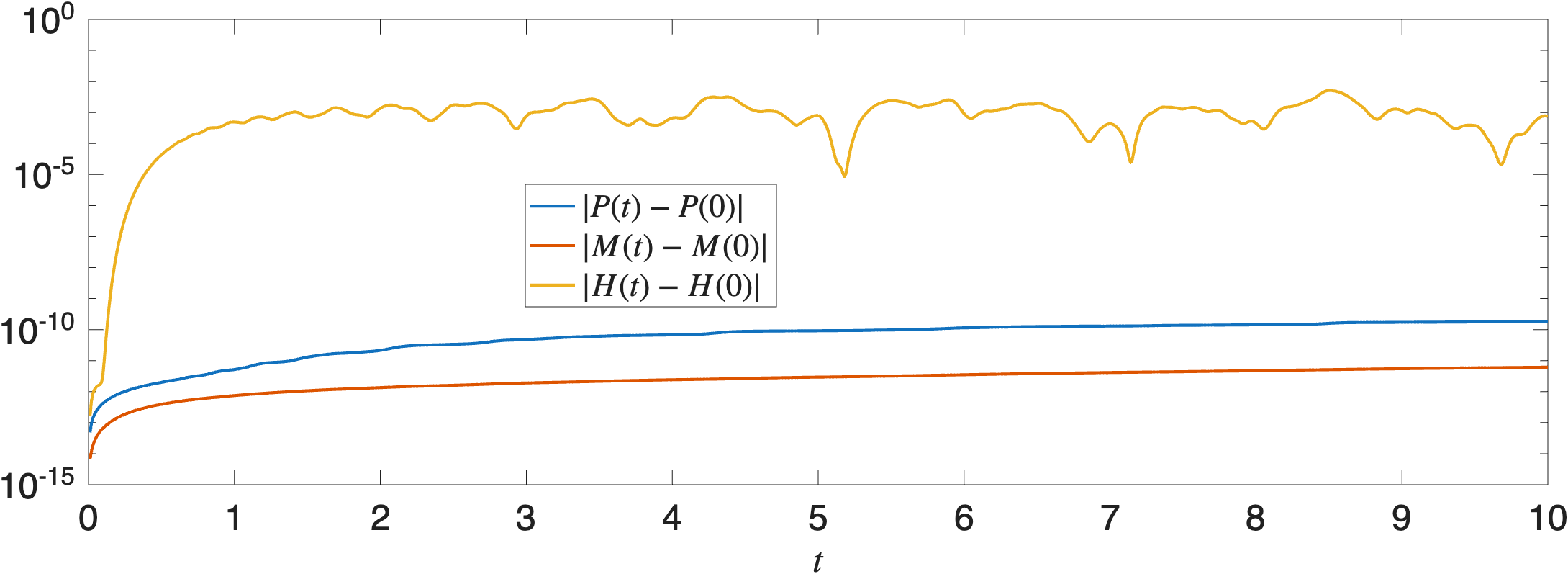}
   \caption{Absolute difference of conserved quantities given in Eq.~(\ref{conserved_quant}), between the initial $t_0 = 0$ and  $t>0$. The discrete approximations 
   are given in Eqs.~(\ref{parseval}), (\ref{discrete_momentum}), and (\ref{discrete_Hamil}), respectively. 
   The computational parameters taken
   are $N=300$ and $\Delta t=10^{-3}$. 
   \label{con_quant}} 
   \end{figure}

To test the performance of the Ch-IF method we consider the evolution corresponding to the localized initial condition
$u(x,0)= \frac{1}{1 + x^2} .  $
The error in the power, momentum, and Hamiltonian over the 
time interval $[0,10]$ is shown in Fig.~\ref{con_quant}. 
The error in each of the quantities is observed to remain bounded over this interval. 
Other localized initial conditions, e.g. Gaussian, exhibit similar results.

\subsection{Modified Korteweg-de Vries (mKdV) equation}


\begin{figure}
    \centering
   \includegraphics[scale=0.3]{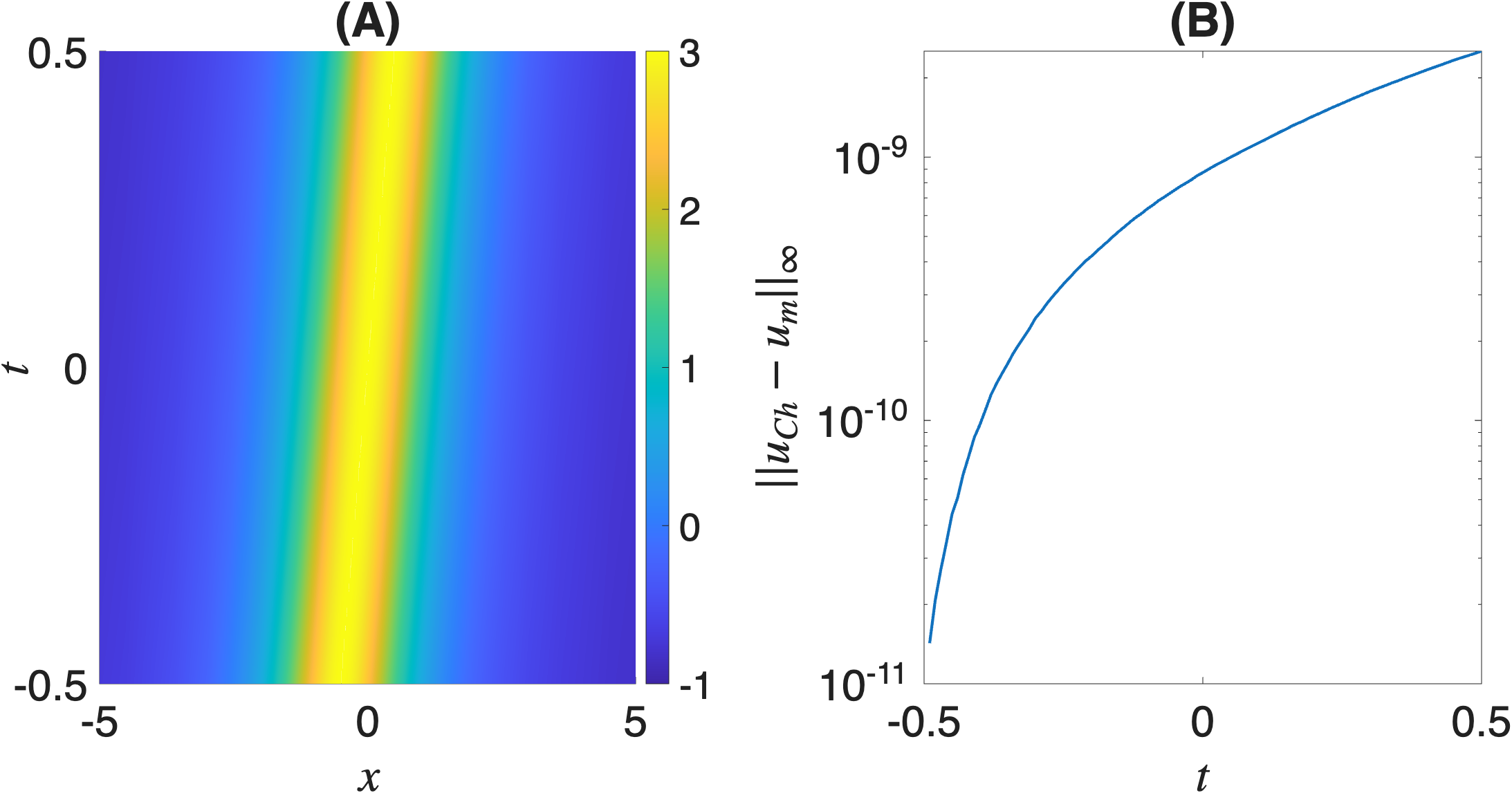}
   \caption{(A) Evolution and (B) error of of the numerical approximation of $u(x,t)$ corresponding to the to the rational soliton (\ref{rational_mkdv}) in the mKdV equation (\ref{mkdv_eq}) for  $t_0=-0.5$. The Ch-IF method is used. Computational parameters are $-0.5\leq t \leq0.5$,
   $N = 114$ 
   and 
   $\Delta t= 10^{-4}.$ 
   \label{mkdv_graph} } 
   \end{figure}

   Next we consider the  modified Korteweg-de Vries (mKdV) equation 
\begin{equation}
    \label{mkdv_eq}
    u_t+u^2u_x+u_{xxx}=0 ,  ~~~~~ u_0 (x) = u(x,t_0) , ~~~~~ (x,t) \in \mathbb{R} \times [t_0,T] .
\end{equation}
The mKdV equation is an integrable PDE \cite{Horikis2025} that has been derived in the context of fluid mechanics \cite{Miura1968,Helal2002}, plasmas \cite{Verheest2016,Verheest2019}, and certain nonlinear optics \cite{Leblond2009,Leblond2011} applications, among others. Furthermore, through the Miura transformation, it is possible to directly relate complex solution of the KdV equation to the mKdV variant \cite{Ablowitz1981}. 

 Below we consider the rational traveling wave solution \cite{Akhmediev2017}
\begin{equation}
    \label{rational_mkdv}
    u_m(x,t)=-1+\frac{12}{3+2(x-t)^2} . 
\end{equation}
 This is a one-hump localized solution with speed 1 and nonzero boundary conditions, $u_m \rightarrow  -1$ as $x \rightarrow \pm \infty$.
To numerically approximate the nonlinear term in Eq.~(\ref{mkdv_eq}), we  express it as $\frac{1}{3} (u^3)_x$, where the nonlinearity is naturally evaluated in the physical domain and the derivative in 
spectral space. Following the algorithm in Appendix~\ref{IVPs}, 
$$\mathbb{L}=-\mathbb{D}_3,~~\mathbb{N}_R[u]=\frac{1}{3}u^3,~~\mathbb{N}_{\text{Ch}}=-\mathbb{D}_1,~~\text{and}~~ U=-1, $$
where $\mathbb{D}_3$ is the third differentiation matrix defined in \ref{third_Ch_matrix} and $\mathbb{D}_1$ is the first differentiation matrix defined in Section \ref{diff_mat_sec}.

\subsubsection{A convergence study}

\begin{figure}
    \centering
   \includegraphics[scale=0.325]{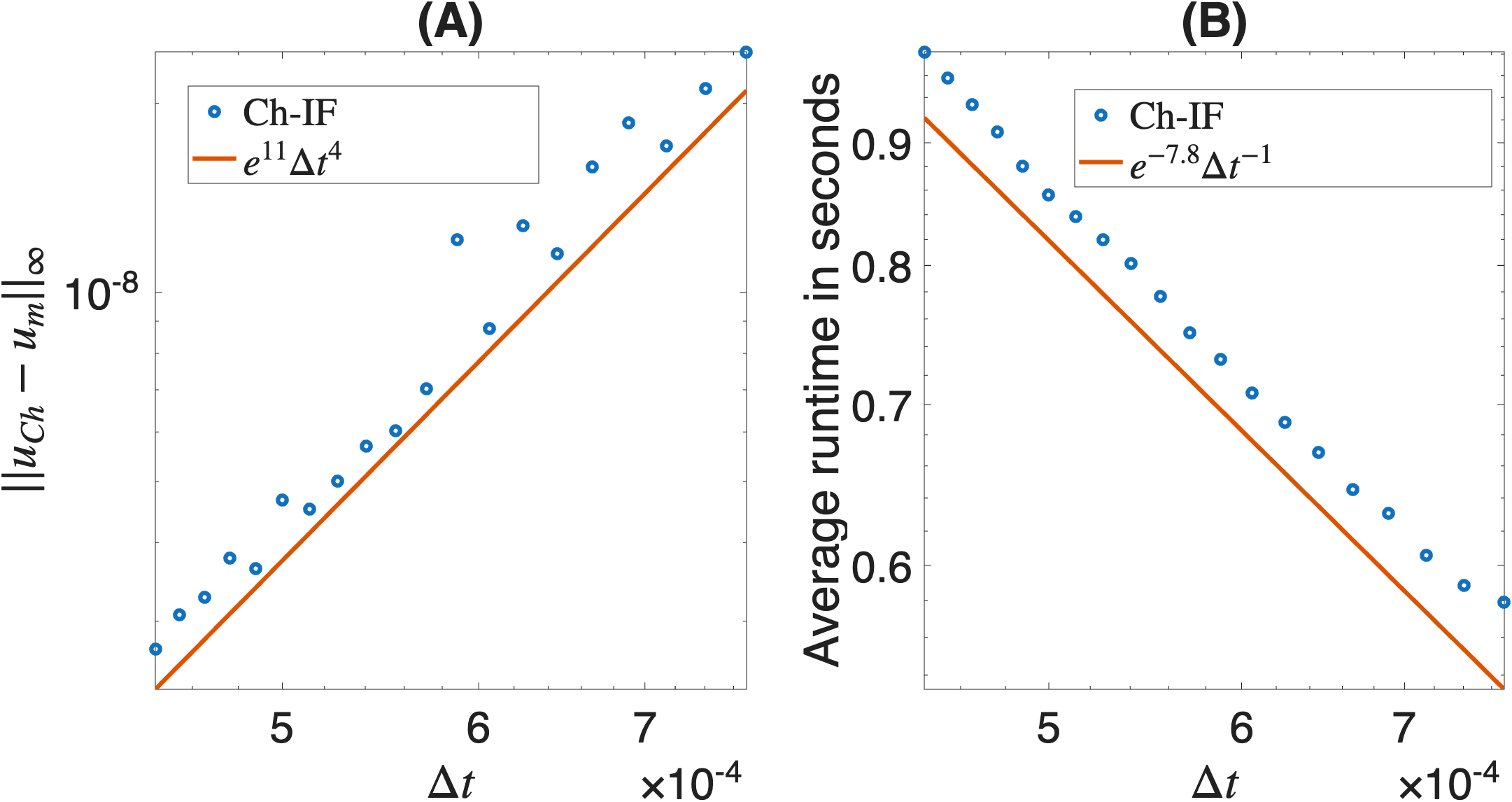}
   \caption{(A) Convergence of 
   Ch-IF approximation for the rational soliton (\ref{rational_mkdv}) solved on the interval $-0.5 \le t \le 0.5$. The global space-time error for $N = 114$  is shown. The Ch-IF method converges  fourth-order as  $\Delta t $ decreases.   (B) Average CPU runtime as a function of $\Delta t$ and fixed $N=114$. The runtime  increases linearly with $\Delta t^{-1}$.
   \label{mkdv_converge}} 
   \end{figure}

Typical results obtained by  solving the mKdV equation (\ref{mkdv_eq})  for the rational soliton solution (\ref{rational_mkdv}) with the Ch-IF method are discussed next. The evolution of a typical traveling soliton obtained using this method is shown in Fig.~\ref{mkdv_graph}(A) over the time interval $[-0.5,0.5]$. In general, we observe it necessary to use a relatively small time-step (compared to NLS and BO equations) for numerical stability. This is to be expected due to the  third-order linear effects. 

 Global spatial error as a function of time is highlighted in Fig.~\ref{mkdv_graph}(B). For a time-step size of $\Delta t = 10^{-4}$, the difference between the exact and numerical approximation is $\mathcal{O}(10^{-7})$ or better. We note that we chose the time frame to be symmetric about the origin. Thus ensuring that the peak of the soliton remains close to the origin. When the peak of the solution moves away from the origin, the error increases dramatically. This is expected based on the Ch modal decay discussion in Section~\ref{decay_sect}; see Eq.~(\ref{decay_funk}). Experience shows that keeping the nontrivial parts of the solution near the origin is important for maintaining high-order accuracy.

Next, the  convergence rate and runtime as a function of $\Delta t$ are discussed. In Fig.~\ref{mkdv_converge}(A) the convergence rate of the Ch-IF method applied to  the rational solution (\ref{rational_mkdv}) is shown. An overall fourth-order convergence rate is observed until other errors become significant. 
Corresponding runtimes are shown in Fig.~\ref{mkdv_converge}
(B). 
We observe a roughly linear dependence for this range of computational parameters. For all runs considered in Fig.~\ref{mkdv_converge}, the total CPU time ranged from 0.5-1 seconds for errors in the range of $O(10^{-9})-O(10^{-8})$.

\subsubsection{Conserved quantities}
\label{mkdv_cq_sect}

For 
localized solutions of the mKdV equation (\ref{mkdv_eq}) that decay to zero, 
the integrals 
\begin{equation}
\label{mkdv_conserved_quant}
P = \int_{-\infty}^{\infty} u^2~ dx , ~~~~~~  H = \int_{-\infty}^{\infty} \left(6u_x^2 - u^4 \right) dx ,
\end{equation}
denoting momentum and energy, respectively, are independent of time. These integrals are numerically approximated in a manner similar to that discussed in Sec.~\ref{conserved_quant_sec}, but for real-valued functions. 

\begin{figure}
    \centering
   \includegraphics[scale=0.325]{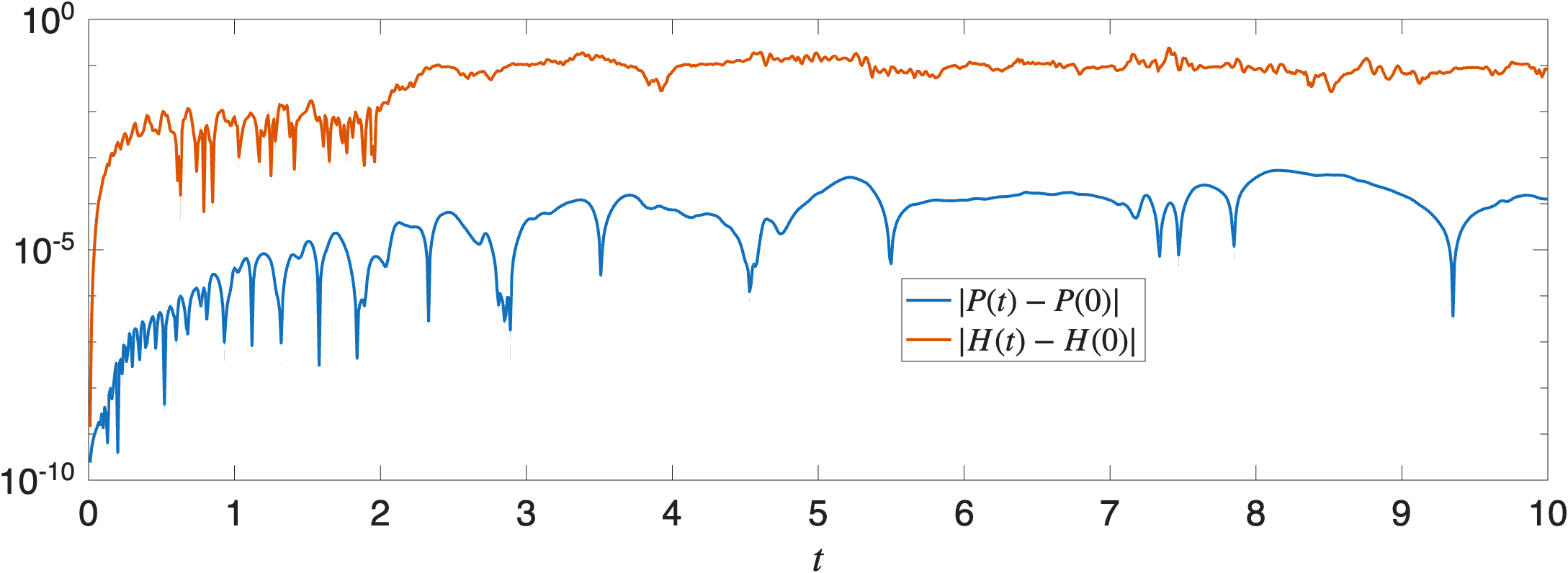}
   \caption{Absolute difference of conserved quantities momentum and energy in Eq.~(\ref{mkdv_conserved_quant}) for initial condition  $u_0(x) =  (1 + x^2)^{-1}$.
   The discrete approximations 
   are given in Eqs.~(\ref{parseval}) and (\ref{discrete_Hamil}), respectively. 
   The computational parameters used are $N=500$
   and $\Delta t = 10^{-3} .$  
   \label{mkdv_con_quant}} 
   \end{figure}

To test the performance of the Ch-IF method we again consider the evolution  
of the localized initial condition 
$u(x,0)= \frac{1}{1 + x^2}.$ The error in the power and Hamiltonian over the 
time interval 
$[0,10]$ is shown in Fig.~\ref{mkdv_con_quant}. 
After some initial growth, both quantities remain bounded. Other localized initial conditions, e.g. hyperbolic secant, give similar results.

\subsection{Benjamin-Ono (BO) equation}
\label{BO_sec}


\begin{figure}
    \centering
   \includegraphics[scale=0.325]{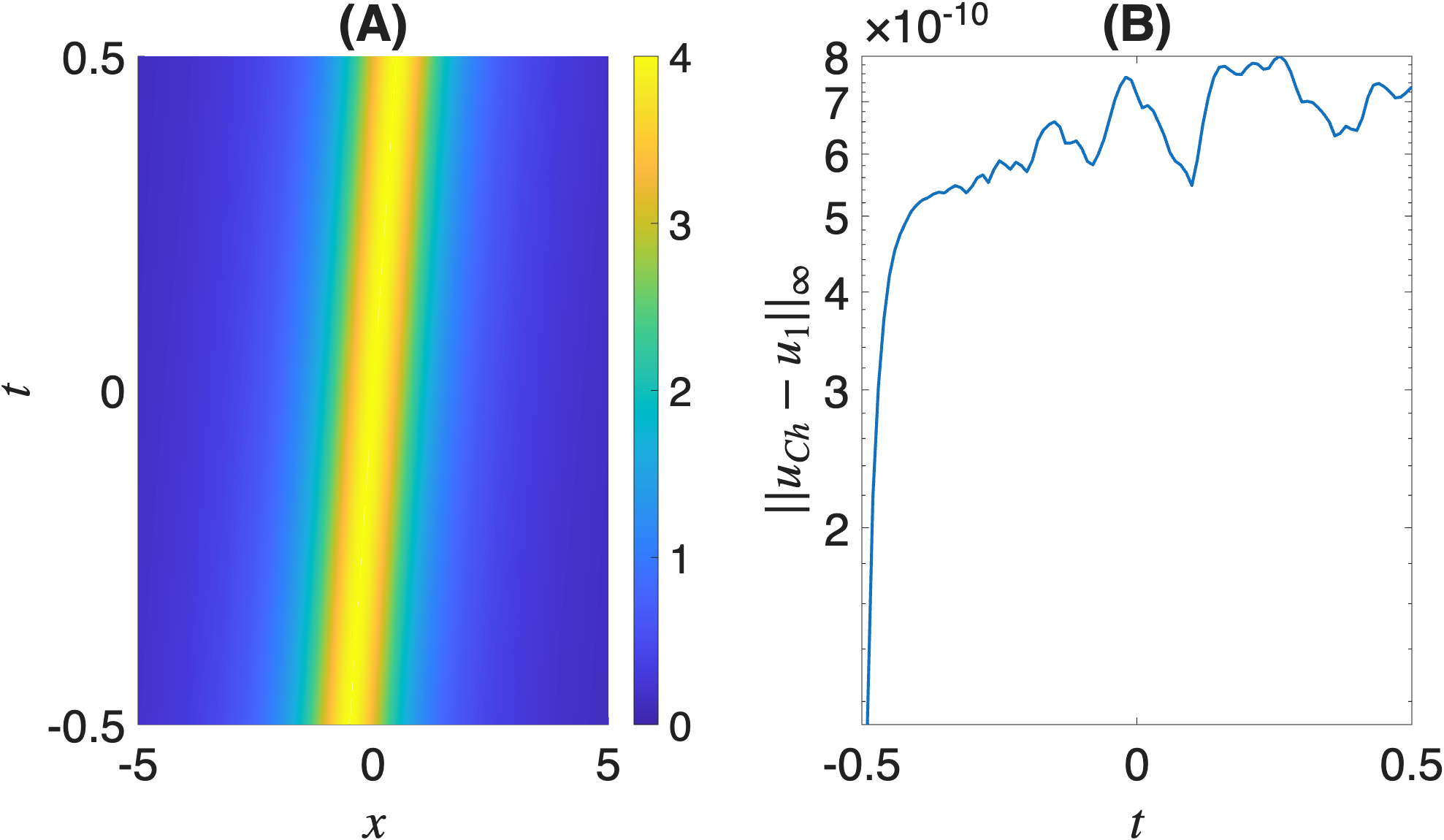}
   \caption{(A) Evolution and (B) error of of the Ch-IF numerical approximation of $u(x,t)$ corresponding to the one soliton solution (\ref{onehump}) in the BO equation (\ref{bo_eq}) for  $t_0=-0.5$ and $c=1$. 
   Computational parameters are: $-0.5\leq t \leq0.5$,
   $N = 100$ 
   and 
   $\Delta t= 10^{-3}$. 
   \label{bo_graph} } 
   \end{figure}

The Benjamin-Ono equation is a nonlinear integrable equation and another worthy candidate for gauging the performance of this numerical scheme. The BO equation was originally derived to describe internal gravity waves in  a stratified fluid \cite{Benjamin1967,Ono1975}.
Consider the BO equation  
\begin{equation}
    \label{bo_eq}
    u_t+uu_{x}+\mathcal{H}[u_{xx}](x)=0 ,  ~~~~~ u_0 (x) = u(x,t_0) , ~~~~ (x,t) \in \mathbb{R} \times [t_0,T] , 
\end{equation}
where $H$ is the Hilbert transform defined in Eq.~(\ref{Hil_tran}). In \cite{Weideman1992,Boyd2011}, the authors used a Christov basis to approximate solutions to the BO equation, however those works do not implement an integrating factor method like we do here. A traveling rational one  soliton solution is given by \cite{Weideman1992}
\begin{equation}
    \label{onehump}
    u_{1}(x,t)=\frac{4c}{c^2(x-ct)^2+1} , 
\end{equation}
for wave speed $c>0$ (see Fig.~\ref{bo_graph}(A)), and a two-soliton solution 
\begin{equation}
    \label{twohump}
u_2(x,t)=\frac{4c_1c_2(c_1\theta_1^2+c_2\theta_2^2+(c_1+c_2)^3c_1^{-1}c_2^{-1}(c_1-c_2)^{-2})}{(c_1c_2\theta_1\theta_2-(c_1+c_2)^2(c_1-c_2)^{-2})^2+(c_1\theta_1+c_2\theta_2)^2} , 
\end{equation}
where $\theta_1=x-c_1t-\varphi_1$, $\theta_2=x-c_2t-\varphi_2$ and $c_1,c_2,\varphi_1,\varphi_2$ are arbitrary constants. Both of these solutions are 
in $L_2(\mathbb{R})$ with zero boundary conditions ($U = 0$).

To numerically solve Eq.~(\ref{bo_eq}), we note that the nonlinear part can be expressed as the derivative of $\frac{1}{2}u^2$, and thus 
the algorithm given in Appendix~\ref{IVPs} takes the values 
$$\mathbb{L}=-\mathbb{H}\mathbb{D}_2,~~\mathbb{N}_R[u]=\frac{1}{2}u^2,~~\mathbb{N}_{\text{Ch}}=-\mathbb{D}_1,~~\text{and}~~ U=0, $$
where $\mathbb{H}$ is the Hilbert transform matrix 
(see Sec.~\ref{diff_mat_sec}), $\mathbb{D}_2$ is the second differentiation matrix 
(see \ref{third_Ch_matrix}),  and $\mathbb{D}_1$ is the first differentiation matrix (see Sec.~\ref{diff_mat_sec}).

\begin{figure}
    \centering
   \includegraphics[scale=0.325]{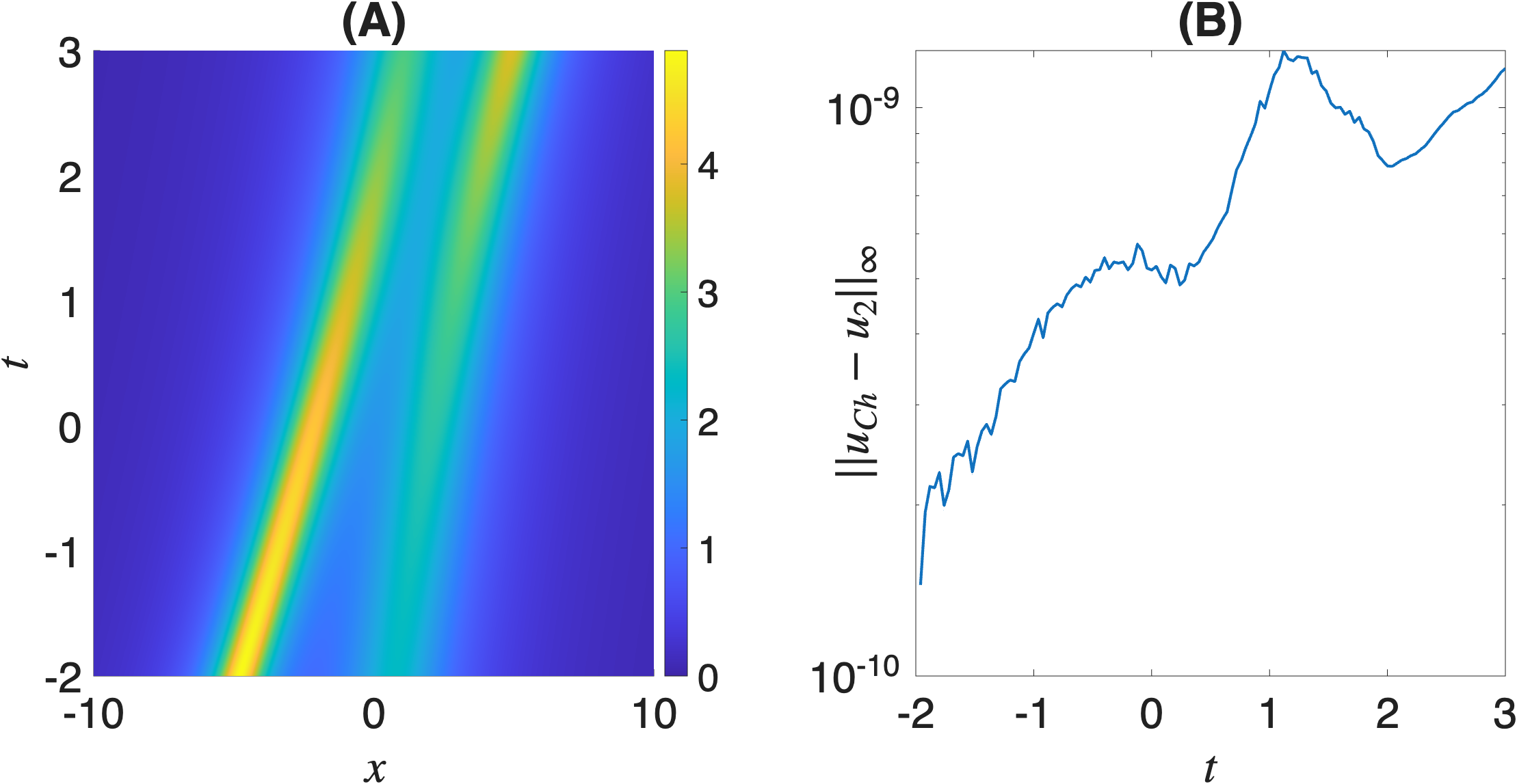}
   \caption{(A) Evolution and (B) error of of the Ch-IF numerical approximation of $u(x,t)$ corresponding to the two-soliton solution (\ref{twohump}) in the BO equation (\ref{bo_eq}) for  $c_1=\frac{3}{2}$, $c_2=\frac{1}{2}$,$\varphi_1=-\frac{1}{2}$ and $\varphi_2=0$ . 
   Computational parameters are: $-2\leq t \leq3$,
   $N = 1600$ 
   and 
   $\Delta t= 10^{-4}$.
   \label{bo_graph2} } 
   \end{figure}

\begin{figure}
    \centering
   \includegraphics[scale=0.325]{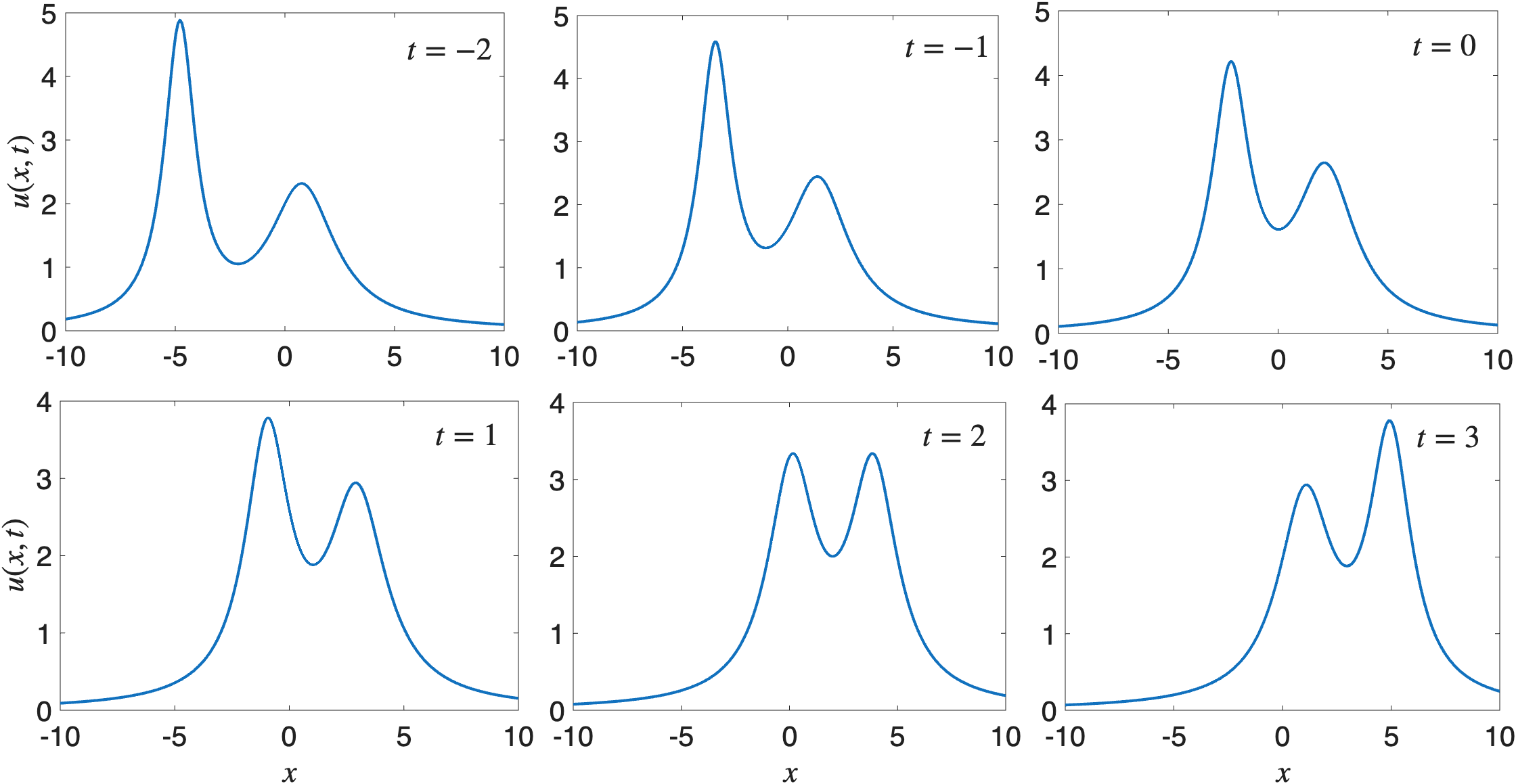}
   \caption{(A) Snapshots $u(x,t)$ corresponding to the two soliton interaction in} Fig.~\ref{bo_graph2}(A). All computational parameters are the same.   \label{bo_snap}  
   \end{figure}

A numerical simulation for the two-soliton solution (\ref{twohump}) is shown in Fig.~\ref{bo_graph2}. A top view of the solution is shown in Fig.~\ref{bo_graph2}(A) over the time interval $[-3,2]$. This panel highlights the interaction of the two solitons and the nonlinear phase shift near time $t \approx 2$.  Notice that the larger (smaller) solitons are moving with a speed of $\frac{3}{2}(1)$. Snapshots of the two solitons interacting are shown in Fig.~\ref{bo_snap}. 
Global spatial error as a function of time is highlighted in Fig.~\ref{bo_graph2}(B). For a time-step size of $\Delta t = 10^{-4}$, the difference between the exact and numerical approximation is $\mathcal{O}(10^{-8})$ or better. In order to adequately resolve this solution, we increase the number Ch modes and consequently must  decrease the time step. This simulation requires more computational resources than the one soliton solution in Fig.~\ref{bo_graph}.

In Fig.~\ref{bo_snap} profile snapshots of the two-soliton interaction in the Benjamin-Ono equation  are shown. These correspond to the solution shown in Fig.~\ref{bo_graph2}.

\subsubsection{A convergence study}

\begin{figure}
    \centering
   \includegraphics[scale=0.325]{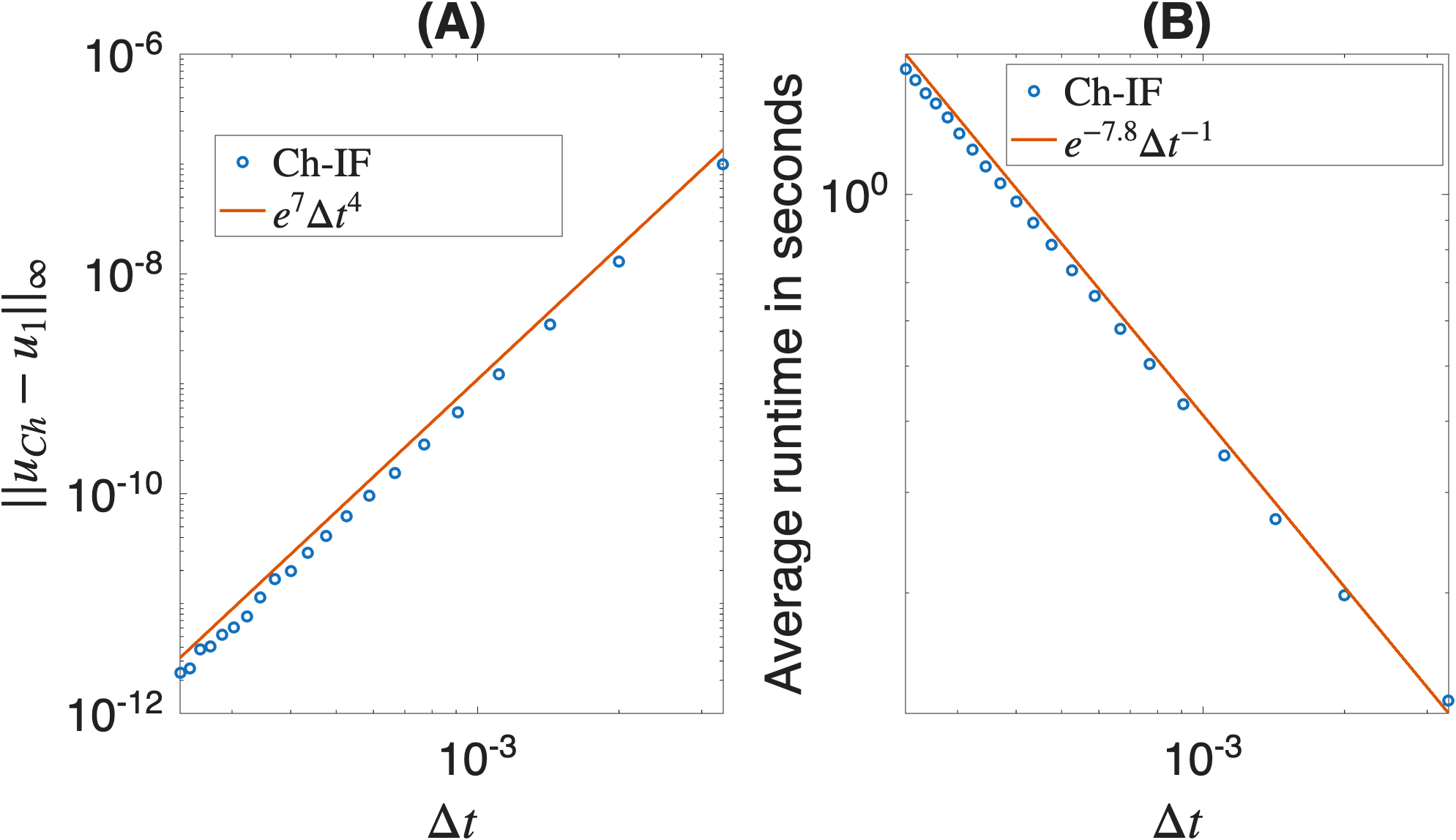}
   \caption{(A) Convergence of 
   Ch-IF approximations for the one soliton solution (\ref{onehump}) solved on the interval $-0.5 \le t \le 0.5$. The global space-time error for $N = 100$ and $c=1$ is shown. The Ch-IF method converges  fourth-order as  $\Delta t $ decreases.   (B) Average CPU runtime as a function of $\Delta t$ and fixed $N=100$. The runtime  increases linearly with $\Delta t^{-1}$. 
   \label{bo_converge}} 
   \end{figure}

Typical results obtained by  solving the BO equation (\ref{bo_eq})  for the one soliton solution (\ref{onehump}) with the Ch-IF method are discussed next. The evolution of a typical one soliton obtained using these methods is shown in Fig.~\ref{bo_graph}(A) over the time interval $[-0.5,0.5]$. Notice that the soliton is moving with a speed of $1$.  Global spatial error as a function of time is highlighted in Fig.~\ref{bo_graph}(B). For a time-step size of $\Delta t = 10^{-3}$, the difference between the exact and numerical approximation is $\mathcal{O}(10^{-9})$ or better. Notably, for this time interval, we do not observe the error growth observed in NLS (see Fig.~\ref{nls_graph}) and mKdV (see Fig.~\ref{mkdv_graph}) equations. This is likely due to the lack of instability in the solution and the solution peak staying relatively close to the origin.

The  convergence rate and runtime as a function of $\Delta t$ are discussed next. In Fig.~\ref{bo_converge}(A) the convergence rate of the Ch-IF method applied to  the one soliton solution (\ref{onehump}) is shown. A  clear fourth-order convergence rate is observed until other errors become significant. 
Corresponding runtimes are shown in Fig.~\ref{bo_converge}(B). 
We observe a roughly linear dependence, similar to the previous examples, for this range of parameters. For all runs considered in Fig.~\ref{bo_converge}, the total CPU time ranged from 0.1-1.3 seconds.

\subsubsection{Conserved quantities}
\label{bo_cq_sect}

\begin{figure}
    \centering
   \includegraphics[scale=0.325]{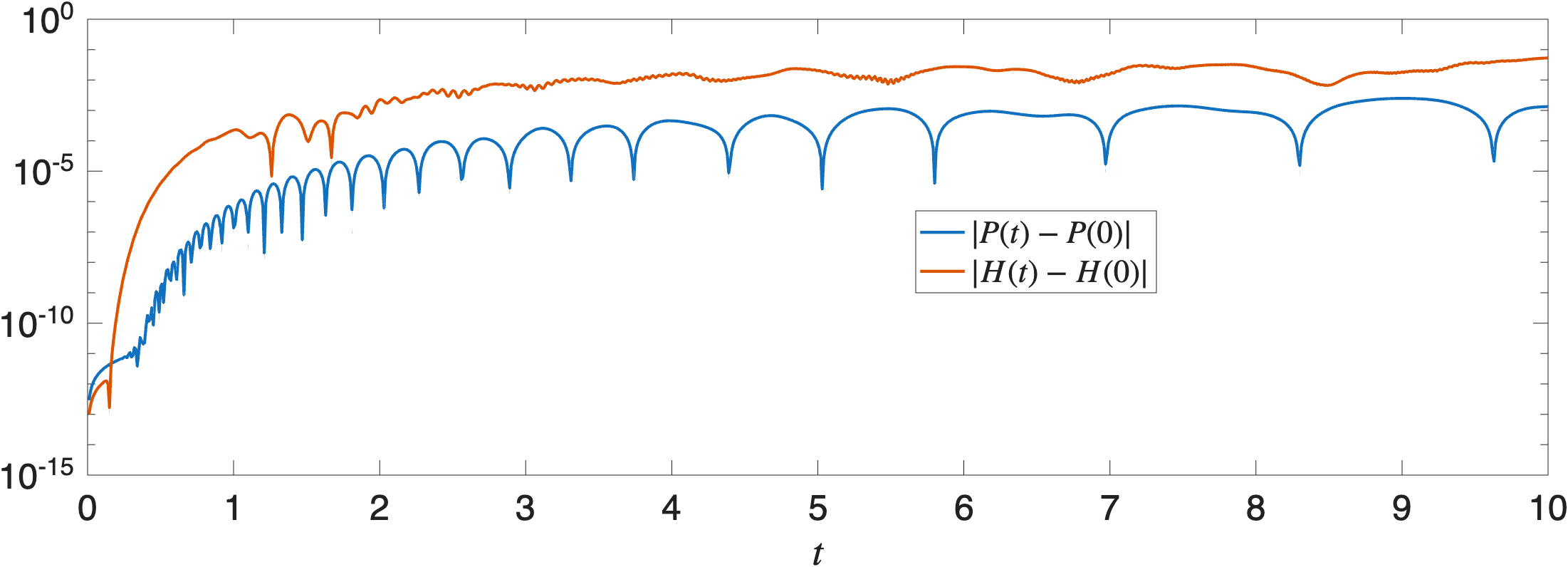}
   \caption{Absolute difference of momentum and energy from Eq.~(\ref{bo_constant}) between the initial $t_0 = 0$ and   at $t>0$. The discrete approximations 
   are given in Eqs.~(\ref{parseval})  and (\ref{discrete_Hamil2}). 
   The computational parameters used are $N=500$ Ch nodes and $\Delta t=10^{-3}$. 
   \label{bo_con_quant}} 
   \end{figure}

For 
localized solutions, 
the BO equation (\ref{bo_eq}) conserves the momentum and energy quantities, respectively, 
\begin{equation}
    \label{bo_constant}
   P = \int_{-\infty}^{\infty} u^2~ dx , ~~~~ H=\int_{-\infty}^{\infty}\left(u^3-3u_x\mathcal{H}[u]\right)~dx .
\end{equation}
Denoting $\rho(x,t) = u^2(x,t)$, the Hamiltonian is numerically evaluated by 
\begin{equation}
\label{discrete_Hamil2}
H  =
\Check{\bm \rho}^\dag \Check{\bf u}  -  3( \mathbb{D}_1  \Check{\bf u})^\dag(\mathbb{H}\check{\bf{u}})  , ~~~~~ \Check{\bm \rho} = (\cdots, \check{\rho}_{-1},  \check{\rho}_{0}, \check{\rho}_{1}, \cdots )^T .
\end{equation}
Momentum is evaluated via the Parseval relation in Eq.~(\ref{parseval}).  For the solutions (\ref{onehump}) and (\ref{twohump}) with zero boundary conditions, the momentum and energy should be conserved and are over the time frame. However, as each of the solutions is traveling the error of the conserved quantities grows with error of the function as the solution moves farther away from the origin. It is for  this reason that we check the conserved quantities with the same initial condition used previously, 
$u(x,0)= \frac{1}{1 + x^2} .  $
The error in the momentum and energy time interval $[0,10]$ is shown in Fig.~\ref{bo_con_quant}. 
After some initial growth, the error in the energy remains bounded. Other localized initial conditions 
exhibit similar results.

\section{Conclusion}
\label{conclude_sec}

In this work an integrating-factor time-stepping method for  evolving rational solutions of nonlinear partial differential equations of physical interest was developed. The novelty lies in the application of an integrating-factor approach with Christov functions, a family 
of orthogonal rational functions defined on the real line. For rational  functions, 
the decay rate of the modal coefficients can be related the location of their singularities in the complex plane. Computationally, this basis is attractive since it can be mapped to a discrete Fourier transform and computed via fast Fourier methods.

An explicit fourth-order Runge-Kutta method was used to time-step the \\integrating-factor method. The method was tested on rational solutions of the nonlinear Schr\"odinger,  modified Korteweg-de Vries, and  Benjamin-Ono equations; all performed well. For the examples shown, we found solutions accurate to $O(10^{-6})$ or better in a few seconds on a standard personal laptop.

Rational solutions appear in many physically important problems, like rogue waves, and motivate the need for an effective basis to accommodate their algebraically slow decay rate. The Christov basis exhibits many attractive qualities for approximating these solutions. 
The biggest limitation of this approach is the clustering of spatial points near the origin. This necessitates the concentration of important dynamics near the origin. For traveling wave solutions, as an example, this will make this a challenging basis to work with (outside the co-moving frame). A potential solution to this problem is an adaptive grid that moves with important wave motion.

\section{Declarations}

\subsection{Conflict of Interest}

The authors declare that they have no conflict of interest.

\subsection{Funding statement} No funding was received or utilized for the research or authorship of this paper.

\subsection{Author Contributions} Both Troy I. Johnson and Justin T. Cole contributed to the conceptualization of this manuscript. The figures were made by Troy I. Johnson.  Both Troy I. Johnson and Justin T. Cole worked on every draft of the manuscript and read and approved the final manuscript.

\subsection{Acknowledgments}
The authors acknowledge motivating and insightful discussions with Denis Silantyev.

\appendix

\section{Decay rate of Ch coefficients continued}
\label{decay_app}

 \begin{figure}
    \centering
   \includegraphics[scale=0.32525]{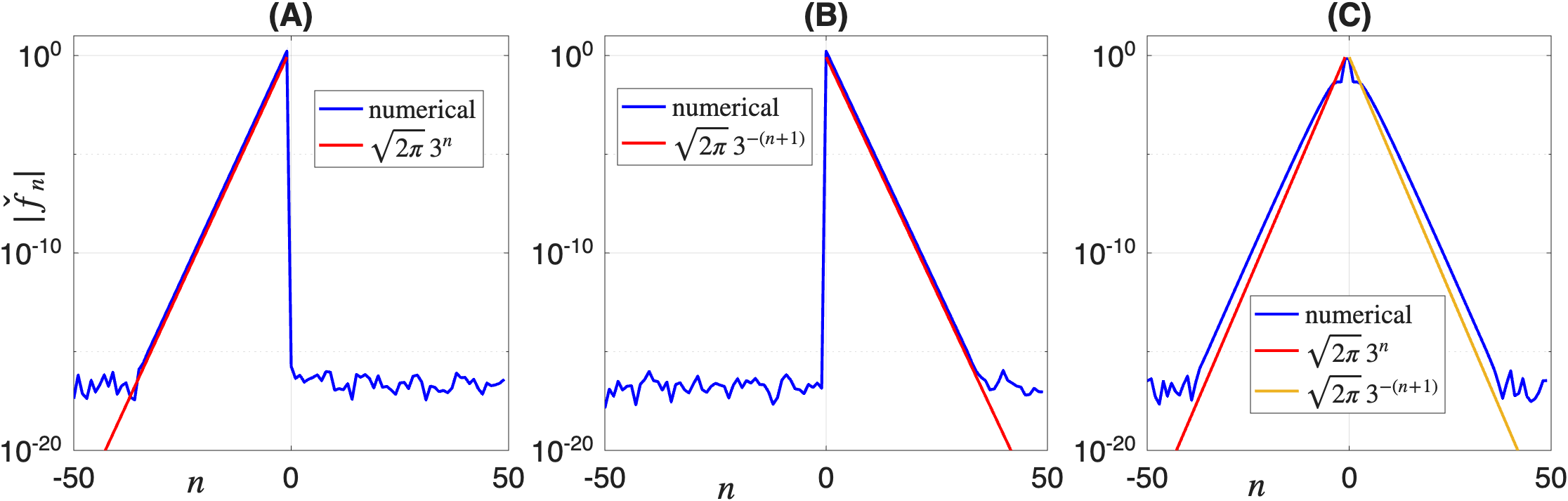}
   \caption{(A) Numerically computed Ch coefficients for the function $f_1(x) = (1+ix)^{-1}$ in blue, the Ch coefficients from (\ref{rational_expand_exact}) are in red. (B) Numerically computed Ch coefficients for the function $f_2(x) = (1-ix)^{-1}$ in blue, the Ch coefficients from  (\ref{rational_expand_exact}) are in red. (C) Numerically computed Ch coefficients for the function $f_3(x) = (1+x^2)^{2}$ in blue, the Ch coefficients from  (\ref{rational_expand_exact}) are in red(yellow). Notice that 
   the Ch coefficients decay exponentially fast. \label{Ch_rcoeff}}
   \end{figure}

In this section we continue the discussion given in Section \ref{decay_sect}. As a second illuminating example, we examine two functions with poles in only in one half plane: upper or lower.
Consider the functions 
\begin{equation}
f_1(x) = \frac{1}{1 + i x}  ~~~~\text{and} ~~~~ f_2(x) = \frac{1}{1 - i x} ,
\end{equation}
which have poles at $i$ and $-i$, respectively. Unlike the previous example, these functions are entire in the lower and upper half planes, respectively. The distance from the real line to the poles in $\theta$, given in Eq.~(\ref{decay_rate}), is $\kappa = \log(3)$. When we compare the decay rates with the numerically computed values in  Fig.~\ref{Ch_rcoeff}(A)-(B), we observe that $(1+ix)^{-1}$ has nontrivial Ch coefficients for nonpositive $n$ and that $(1-ix)^{-1}$ has nontrivial Ch coefficients for nonnegative $n$. We  note that $(1+ix)^{-1}(1-ix)^{-1}=(1+x^2)^{-1}$, and the decay rates of the nontrivial numerical Ch coefficients of $f_1$ and $f_2$ match those found in (\ref{rational_expand_exact}) for the negative and positive modes, respectively.

To understand this asymmetric behavior, return to the Fourier-Ch coefficients in Eq.~(\ref{Ch_coeff_theta}) and take the complexification of $\theta = \psi + i \phi$: $e^{-i n \theta} = e^{- i n \psi} e^{ n \phi}$. For $f_1$, an analytic strip can be extended up to the pole at $\kappa = \log(3) ~(z = i)$, and so for $n < 0$, the exponential decay is of the form $e^{-|n| \log (3)} = 3^{-|n|}$. On the other hand, there is no lower bound on how far the analytic strip may be extended in the lower half plane, and $\phi \rightarrow - \infty$. Hence, for $n > 0$ the decay rate is faster than any exponential and we observe a compact support in Fig.~\ref{Ch_rcoeff}(A). A similar argument follows for $f_2$, but with the behavior in the upper and lower half planes reversed (see Fig.~\ref{Ch_rcoeff}(B)).

As a final example we consider the case of repeated roots 
$$f_3(x)=(1+x^2)^{-2}$$ which has double poles at $z=\pm i$. When we compare the decay rates with the numerically computed values in  Fig.~\ref{Ch_rcoeff}(C) we see that the  decay rate is similar  to those in Eq.~(\ref{rational_expand_exact}). Thus, the multiplicity of a pole has a nominal effect on the decay rate of a functions Ch coefficients with the main factor being the distance from nearest pole to the real axis in $\theta$.

\section{Higher order Christov differentiation matrices}
\label{third_Ch_matrix}

In this section we give the Ch differentiation matrices for the second and third 
derivative. The process for constructing them resembles that used to find the first derivative matrix in Eq.~(\ref{deriv_coeff_relate}). 

 
 Consider a function $f$ expressed in basis of Ch functions, i.e. (\ref{Ch_expansion}), whose  second derivative is clearly
 \begin{equation*}
\label{Ch_expansion_2ndDerivative}
 f''(x) = \sum_{n = - \infty}^{\infty} \Check{f}_n\phi''_n(x) = \sum_{n = - \infty}^{\infty} \Check{f}^{(2)}_n \phi_n(x) ,
 \end{equation*}
where $\Check{f}^{(2)}_n$ denotes the second derivative coefficients, in terms of a basis of Ch functions. Differentiating the derivative relation in Eq.~(\ref{deriv_recurrence}) and shifting indices \cite{Cole2025}, we obtain the second derivative coefficients are related to the function coefficients via
\begin{equation}
\label{2ndderiv_coeff_relate}
\Check{f}^{(2)}_n = n(n-1)  \Check{f}_{n-2} + 4in^2 \Check{f}_{n-1}  -(2+6n(n+1))  \Check{f}_{n}  -4i(n+1)^2 \Check{f}_{n+1} + (n+2) (n+1)  \Check{f}_{n+2}   .
\end{equation}
The Ch coefficients of the second derivative can also be computed via matrix multiplication
\begin{equation}
\label{Ch_2ndderiv_form}
\check{\textbf{f}}^{(2)} = \left( \dots , \check{f}^{(2)}_{-1}, \check{f}^{(2)}_0, \check{f}^{(2)}_1 , \dots \right)^T, ~~~~ \check{\textbf{f}}^{(2)} =\mathbb{D}_2 \check{\textbf{f}} , ~~~~ \check{\textbf{f}} = \left( \dots , \check{f}_{-1}, \check{f}_0, \check{f}_1 , \dots \right)^T ,
\end{equation} 
where $\mathbb{D}_2$ is the  banded  matrix

\setlength{\arraycolsep}{1.5 pt}
$$
\mathbb{D}_2 =
\left[
\begin{array}{cccccccccccccccc}
\ddots & \ddots & \ddots & \ddots & \ddots \\
       & 30     & 100i   & -122   & 64i    & 12 \\
       &        & 20     & 64i    & -74    & 36i   & 6 \\
       &        &        & 12     & 36i    & -38   & 16i   & 2 \\
       &        &        &        & 6      & 16i   & -14   & -4i   & 0 \\
       &        &        &        &        & 2     & 4i    & -2    & 0     & 0 \\
       &        &        &        &        &       & 0     & 0     & -2    & -4i   & 2 \\
       &        &        &        &        &       &       & 0     & 4i    & -14   & -16i  & 6 \\
       &        &        &        &        &       &       &       & 2     & 16i   & -38   & -36i  & 12 \\
       &        &        &        &        &       &       &       &       & 6     & 36i   & -74   & -64i  & 20 \\
       &        &        &        &        &       &       &       &       &       & 12    & 64i   & -122  & -100i & 30 \\
       &        &        &        &        &       &       &       &       &       &       & \ddots & \ddots & \ddots & \ddots
\end{array}
\right].
$$
This  matrix is pentadiagonal, Hermitian ($\mathbb{D}_2^\dag = \mathbb{D}_2$), and quite sparse.

Next, we seek to approximate the third derivative by
 \begin{equation}
 f'''(x)  = \sum_{n = - \infty}^{\infty} \Check{f}_n \phi'''_n(x)  = \sum_{n = - \infty}^{\infty} \Check{f}^{(3)}_n \phi_n(x)   ,
 \end{equation}
where $\Check{f}^{(3)}_n$ denotes Ch coefficients of the third derivative function for an expansion in terms of
Ch functions. Differentiating the derivative relation in Eq.~(\ref{deriv_recurrence}) twice and shifting indices yields the relation 

\begin{align}
\check{f}^{(3)}_n  &= (n-2) (n-1) n \check{f}_{n-3} + 3 i (n-1)n (2n - 1) \check{f}_{n-2} - 3 n (5 (n^2 - 1)+ 6)  \check{f}_{n-1}\\\nonumber
& -2i(2n+1)(5n(n+1)+3) \check{f}_n + 3 (n+1)(5(n+1)^2+1) \check{f}_{n+1}\\ \nonumber
& + 3 i (n+1) (n+2) (2n + 3)\check{f}_{n+2} -(n+3)(n+2)(n+1) \check{f}_{n+3} ,
\end{align}
where there is clearly dependence on the seven nearest Ch modes.

Similar to the previous cases, the coefficients of the third derivative can be computed via
\begin{equation}
\label{Ch_3ndderiv_form}
\check{\textbf{f}}^{(3)} = \left( \dots , \check{f}^{(3)}_{-1}, \check{f}^{(3)}_0, \check{f}^{(3)}_1 , \dots \right)^T, ~~~~ \check{\textbf{f}}^{(3)} =\mathbb{D}_3 \check{\textbf{f}} , ~~~~ \check{\textbf{f}} = \left( \dots , \check{f}_{-1}, \check{f}_0, \check{f}_1 , \dots \right)^T ,
\end{equation} 
where $\mathbb{D}_3$ is  the 
banded  matrix

\setlength{\arraycolsep}{1 pt}
\scalebox{0.50}{$\mathbb{D}_3 =   \left[
\begin{array}{*{24}c}
\ddots & \ddots & \ddots & \ddots & \ddots & \ddots & \ddots &  &  &  &  &  &  &  &  &  &  &  &  &  &  &  &  & \\
 & -720 & -3672i & 7704 & 8490i & -5166 & -1638i & 210 &  &  &  &  &  &  &  &  &  &  &  &  &  &  &  & \\
 &  & -504 & -2520i & 5166 & 5538i & -3258 & -990i & 120 &  &  &  &  &  &  &  &  &  &  &  &  &  &  & \\
 &  &  & -336 & -1638i & 3258 & 3366i & -1890 & -540i & 60 &  &  &  &  &  &  &  &  &  &  &  &  &  & \\
  &  &  &  & -210 & -990i & 1890 & 1854i & -972 & -252i & 24 &  &  &  &  &  &  &  &  &  &  &  &  & \\
   &  &  &  &  & -120 & -540i & 972 & 882i & -414 & -90i & 6 &  &  &  &  &  &  &  &  &  &  &  & \\
     &  &  &  &  &  & -60 & -252i & 414 & 330i & -126 & -18i & 0 &  &  &  &  &  &  &  &  &  &  & \\
        &  &  &  &  &  &  & -24 & -90i & 126 & 78i & -18 & 0 & 0 &  &  &  &  &  &  &  &  &  & \\
        &  &  &  &  &  &  &  & -6 & -18i & 18 & 6i & 0 & 0 & 0 &  &  &  &  &  &  &  &  & \\
         &  &  &  &  &  &  &  &  & 0 & 0 & 0 & -6i & 18 & 18i & -6 &  &  &  &  &  &  &  & \\
         &  &  &  &  &  &  &  &  &  & 0 & 0 & -18 & -78i & 126 & 90i & -24 &  &  &  &  &  &  & \\
          &  &  &  &  &  &  &  &  &  &  & 0 & 18i & -126 & -330i & 414 & 252i & -60 &  &  &  &  &  & \\
          &  &  &  &  &  &  &  &  &  &  &  & 6 & 90i & -414 & -882i & 972 & 540i & -120 &  &  &  &  & \\
           &  &  &  &  &  &  &  &  &  &  &  &  & 24 & 252i & -972 & -1854 & 1890 & 990i & -210 &  &  &  & \\
            &  &  &  &  &  &  &  &  &  &  &  &  &  & 60 & 540i & -1890 & -3366i & 3258 & 1638i & -336 &  &  & \\
             &  &  &  &  &  &  &  &  &  &  &  &  &  &  & 120 & 990i & -3258 & -5538i & 5166 & 2520i & -504 &  & \\
              &  &  &  &  &  &  &  &  &  &  &  &  &  &  &  & 210 & 1638i & -5166 & -8490i & 7704 & 3672i & -720 & \\
              &  &  &  &  &  &  &  &  &  &  &  &  &  &  &  &  & \ddots & \ddots & \ddots & \ddots & \ddots & \ddots & \ddots 
\end{array}
\right] .$}

This matrix is septadiagonal, skew-Hermitian, and fairly sparse. 

\begin{figure}
\centering
   \includegraphics[scale=0.325]{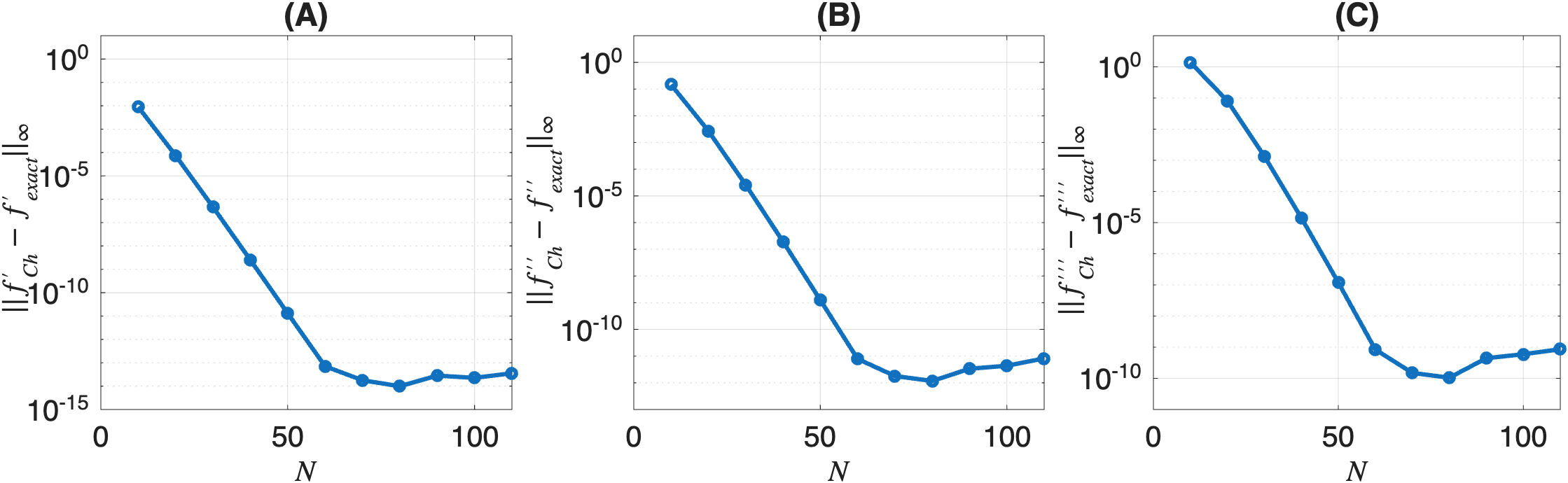}
 \caption{Infinity norm error of the (A) first, (B) second and (C) third Ch derivative approximations applied to the rational function $f(x) = (1 + x^2)^{-1}$ as a function of the number of Ch modes, $N$. The error approaches zero exponentially fast. 
 \label{Deriv_converge}}
   \end{figure}

A typical set of convergence results is highlighted in Fig.~\ref{Deriv_converge}. For the rational function $f(x) = 1/(1 +x^2)$, the Ch approximation of derivatives is remarkably effective. The error converges exponentially fast to zero until we reach round-off error effects at around $N = 70$ modes.  We notice with each order of derivative, the error  increases by about one order of magnitude. These rapid convergence results are expected based on the exponentially decay of the Ch coefficients shown in Fig.~\ref{Ch_coeff_decay_numerical}. Given that we are using fast modified Fourier transforms in (\ref{Ch_Fourier_abbrev}) and (\ref{MIFFT}), these calculations are performed quickly.

\section{Christov integrating-factor method details}
\label{IVPs}

All examples considered in this work are solved using the Christov integrating-factor (Ch-IF) method described in Sec.~\ref{num_pde_sec}. This approach uses a spectral Ch approximation in space and a fourth-order Runge-Kutta IF method in time. The algorithm pseudo-code and technical details are given below.

The time interval $[t_0,T]$ is discretized into $N_t+1$ equally spaced time levels $t_k = t_0 + k \Delta t$ for  $k = 0 ,1 , \dots, N_t$ with time-steps of $\Delta t  = \frac{T-t_0}{N_t}$. The real line $(- \infty , \infty)$  is discretized by $x_{j + \frac{N}{2} +1 }=\frac{1}{2}\tan\left( \frac{ \pi j }{N}  \right),$ for $ j = - \frac{N}{2},- \frac{N}{2} +1 ,..., \frac{N}{2} -1$ and  
$ {\bf x}=(x_1,x_2,...,x_{N})^T$ for even $N$; note that $x_1 = - \infty$.   The numerical approximation of the solution is denoted by ${\bf u}(t) =(u_1,u_2,...,u_{N})^T $ where $u_j  \approx u(x_j,t )$. The initial condition (IC) is given by ${\bf u}_0 = u({\bf x}, t_0)$.  We choose $N$ large enough so that the largest Ch coefficient of the IC
$$\Check{w}_N(t_0) =  \langle \phi_N , u(x,t_0)-U   \rangle = \int_{-\infty}^{\infty} \phi^*_n(x) \Big(u(x,t_0)-U \Big)  ~ dx , $$
is sufficiently small, where $w = u - U$. Consider the initial value problem given in Eq.~(\ref{general_nonlin_PDE}) and the subsequent IF form in Eq.~(\ref{general_if_form}); the linear operator  given by
$$\mathbb{L}= \left[ c_1\mathbb{D}_1+c_2\mathbb{D}_2+c_3\mathbb{D}_3+ \cdots \right] + \mathbb{H} \left[ d_1\mathbb{D}_1+d_2\mathbb{D}_2+d_3\mathbb{D}_3+ \cdots \right], $$
with $\mathbb{D}_j$ being the Ch differentiation matrices and the Hilbert transform matrix defined in Section \ref{diff_mat_sec} and Appendix \ref{third_Ch_matrix}, 
now truncated to matrices of size $ N \times N$. The discretized integrating factors for one time step may be pre-computed. 
$$E=\text{exp}[-\mathbb{L}\Delta t/2], ~~~~~~E_2=E^2$$
The nonlinear operator, $\mathcal{N}$, is assumed to depend on some 
product of $u$ and its derivatives or Hilbert transform.  If it depends on the latter, we apply the appropriate operator 
in spectral space while all nonlinearities, like $u^p$, are evaluated in physical space. As such, we break the nonlinear operator into two parts, 
$\mathbb{N}_R[u]$, to be applied in real space and $\mathbb{N}_{\text{Ch}}$, which is a matrix that'll be applied in spectral space (see below). These operators for the examples in Section~\ref{example_Sec} are given.
Below we give the fourth-order Runge-Kutta Integrating Factor method. \\

\noindent
{\it One Ch-IF time-step} 

\begin{align*}
\Check{{\bf v}} =\text{MFFT}[{\bf u}(t_j)-U]\\
{\bf K}_1=\mathbb{N}_{\text{Ch}}\text{MFFT}\Big[\mathbb{N}_R[{\bf u}]-\mathbb{N}_R[U]\Big]\\
{\bf w}=\text{MIFFT}\left[E^{-1}\left(\Check{{\bf v}}+\frac{\Delta t}{2}{\bf K}_1\right)\right]\\
{\bf K}_2=E\Big(\mathbb{N}_{\text{Ch}}\text{MFFT}\Big[\mathbb{N}_R[{\bf w}+U]-\mathbb{N}_R[U]\Big]\Big)\\
{\bf w}=\text{MIFFT}\left[E^{-1}\left(\Check{{\bf v}}+\frac{\Delta t}{2}{\bf K}_2\right)\right]\\
{\bf K}_3=E\Big(\mathbb{N}_{\text{Ch}}\text{MFFT}\Big[\mathbb{N}_R[{\bf w}+U]-\mathbb{N}_R[U]\Big]\Big)\\
{\bf w}=\text{MIFFT}\left[E^{-1}_2\left(\Check{{\bf v}}+\Delta t{\bf K}_3\right)\right]\\
{\bf K}_4=E_2\Big(\mathbb{N}_{\text{Ch}}\text{MFFT}\Big[\mathbb{N}_R[{\bf w}+U]-\mathbb{N}_R[U]\Big]\Big)\\
\Check{{\bf v}}=\Check{{\bf v}}+\left({\bf K}_1+2\left({\bf K}_2+{\bf K}_3\right)+{\bf K}_4\right)/6\\
{\bf u}{(t_{j+1})}=\text{MIFFT}\left[E_2^{-1}\Check{{\bf v}}\right]+U\\
 t_{j+1} =t_j+\Delta t
\end{align*}

\section{Comparison to Split-Step}
\label{splitstep}
In \cite{Cole2025b} we developed three different fourth-order Christov split-step methods. Namely, we examined the Yoshida, partitioned Runge-Kutta (PRK), and Runge-Kutta-Nystr\"om (RKN) split-step methods. For the NLS equation, we found the PRK and RKN methods to be superior to Yoshida.

\begin{figure}
    \centering
   \includegraphics[scale=0.3]{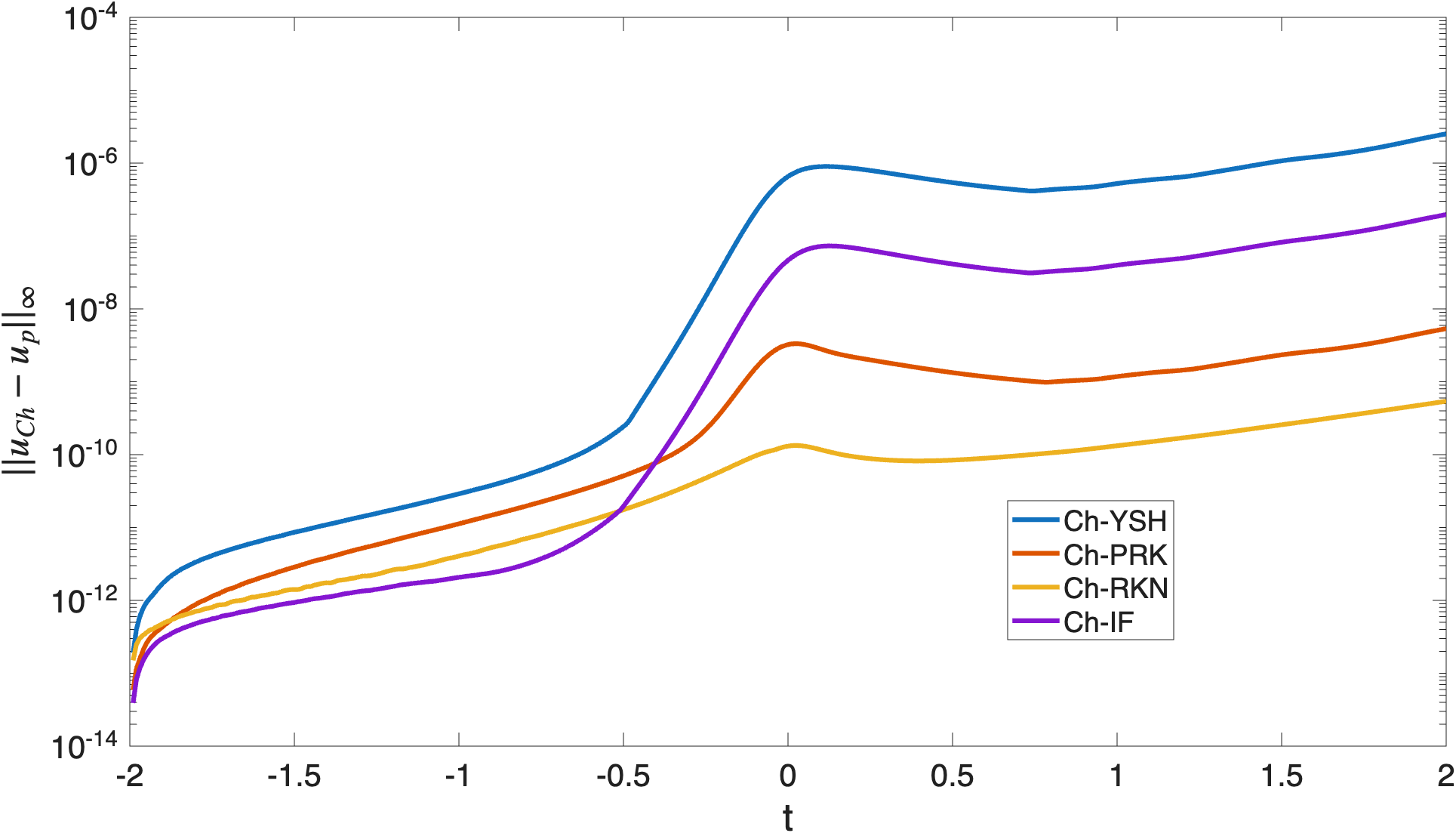}
   \caption{Error in numerical approximation of Peregrine soliton (\ref{peregrine}) over time interval $[-2,2]$ using the Ch split-step methods discussed in \cite{Cole2025b} and the Ch-IF method discussed previously for the NLS equation (\ref{nls_eq}) with $t_0=-2$. The computational parameters   are $N = 300$  and $\Delta t = 0.001 $. 
   \label{nls_comp}} 
   \end{figure}

We compare these methods to Ch-IF method for the same parameters in Fig.~\ref{nls_comp}. 
The Ch-IF methods is comparable to the split-step methods: more accurate than Yoshida, but not quite as accurate as PRK and RKN. As such, for the NLS equation in particular, we view a split-step scheme as more accurate for comparable parameters. However, while the split-step method is well suited for NLS-type equations, where the nonlinear part is simple to solve, they will struggle with the mKdV and BO equations explored in the main article. The advantage of the IF approach is its versatility: accommodating the different nonlinearities is straightforward within our framework.

\bibliographystyle{siamplain}

\begin{thebibliography}{00}
\bibitem{Fornberg1996} B. Fornberg, {\em A practical guide to pseudospectral methods}, Cambridge University Press (1996).
\bibitem{Canuto2006} C. Canuto, M. Y. Hussaini, A. Quarteroni, and T. A. Zang, {\em Spectral Methods: Fundamentals in Single Domains}, Springer-Verlag, Berlin (2006).

\bibitem{Yang2010} J. Yang, {\em Nonlinear Waves in Integrable and Nonintegrable Systems}, SIAM, Philadelphia (2010).
\bibitem{Grosch1977} C. E. Grosch and S. A. Orszag, {\em  Numerical solution of problems in unbounded regions: coordinate transforms}, J. Comp.  Phys. {\bf 25} pp. 273-296 (1977).
\bibitem{Boyd1987} J. P. Boyd, {\em Spectral methods using rational basis functions on an infinite interval}, J. Comp. Phys. {\bf 69}, pp. 112-142 (1987).
\bibitem{Boyd1980} J. P. Boyd, {\em  The rate of convergence of Hermite function series}, Math. of Comp. {\bf 35} pp. 1309-1316 (1980).
\bibitem{Christov1982} C. Christov, {\em A complete orthonormal system of functions in $l^2(-\infty,\infty)$ space}, SIAM J. Appl. Math. {\bf 42} pp. 1337-1344 (1982).
\bibitem{Boyd1990} J. P. Boyd, {\em The orthogonal rational functions of Higgins and Christov and algebraically mapped Chebyshev polynomials}, J. Approx. Theory {\bf 61} pp. 98-105 (1990).
\bibitem{Iserles2020} A. Iserles and M. Webb, {\em A family of orthogonal rational functions and other orthogonal systems with a skew-Hermitian differentiation matrix}, J. Fourier Anal. and Appl. {\bf 26}, 19 (2020).
\bibitem{Boyd2001} J. P. Boyd, {\em  Chebyshev and Fourier Spectral methods} 2nd ed., Dover, New York,  (2001).
 \bibitem{Paley1934} R. E. A. C. Paley and N. Wiener, {\em  Fourier transforms in the complex domain}, Amer. Math. Soc. Colloq. Publ.,  {\bf 19} Providence,  (1934).


\bibitem{Cole2025} J. T. Cole and T. I. Johnson, {\em A split-step Christov method for approximating rational PDE solutions}, J. Comp.  Phys. (2025).

\bibitem{Kharif2003} C. Kharif and E. Pelinovsky, {\em Physical mechanisms of the rogue wave phenomenon}, European. J. Mechanics-B/Fluids {\bf 22} pp. 603-634 (2003).

\bibitem{Chabchoub2011} A. Chabchoub, N. P Hoffmann, and N. Akhmediev, {\em Rogue wave observation in a water wave tank}, Phys. Rev. Lett. {\bf 106} 204502 (2011).

\bibitem{Agrawal2013} G. Agrawal, {\em Nonlinear fiber optics}, Academic Press, Elsevier (2013).
\bibitem{Higgins1977} J. R. Higgins, {\em Completeness and Basis Properties of Sets of Special Functions}, Cambridge University Press (1977).

\bibitem{Weideman1995} J. A. C. Weideman, {\em Computing the Hilbert transform on the real line}, Math. Comp.  {\bf 64} pp. 745-762 (1995).

\bibitem{Ablowitz2011} M. J. Ablowitz, {\em Nonlinear dispersive waves}, Cambridge Press (2011).

\bibitem{Calini2019} A. Calini, C. M. Schober, and M. Strawn {\em Linear instability of the Peregrine breather: Numerical and
analytical investigations}, Appl. Num. Math. {\bf 141} pp. 36-43 (2019).

\bibitem{Ablowitz2021} M. J. Ablowitz and J. T. Cole, {\em Transverse instability of rogue waves},  Phys. Rev. Lett.  {\bf 127} 104101 (2021).
\bibitem{Yang2012} Y. Ohta and J. Yang, {\em General high-order rogue waves and their dynamics in the nonlinear Schr\"odinger equation}, Proc. Royal Soc. A {\bf 468} pp. 1716-1740 (2012).

\bibitem{Biondini2016} G. Biondini and D. Mantzavinos, {\em Universal nature of the nonlinear stage of modulational instability}, Phys. Rev. Lett. {\bf 116} 043902 (2016).

\bibitem{Yuen1978} H. C. Yuen and W. E. Ferguson, {\em Relationship between Benjamin-Feir instability and recurrence in the nonlinear Schr\"odinger equation}, Phys. Fluids {\bf 21} pp. 1275-1278 (1978).
\bibitem{Horikis2025} S. Baqer, T.P. Horikis and D.J. Frantzeskakis, {\em Physical vs mathematical origin of the extended KdV and mKdV equations}, AIMS Mathematics {\bf 10} pp.9295-9309 (2025).
\bibitem{Miura1968} R.M. Miura, C.S. Gardner and M.D. Kruskal, {\em Kortweg-de Vries equation and generalizations, II: existence of conservation laws and constants of motion}, Journal of Mathematical Physics {\bf 9} pp. 1204-1209 (1968).
\bibitem{Helal2002} M.A. Helal, {\em Soliton solution of some nonlinear partial differential equations and its applications in fluid mechanics}, Chaos, Solitons \& Fractals {\bf 13} pp. 1917-1929 (2002).
\bibitem{Verheest2016} F. Verheest, C.P. Oliver and W.A. Hereman, {\em Modified kortweg-de vries solitons at supercritical densities in two-electron temperature plasmas}, Journal of Plasma Physics {\bf 82} (2016).
\bibitem{Verheest2019} F. Verheest and W.A. Hereman, {\em Collissions of acoustic solitons and their electric fields in plasmas at critical compositions}, Journal of Plasma Physics {\bf 85} (2019).

\bibitem{Leblond2009} H. Leblond and D. Mihalache, {\em Few-optical-cycle soltions: modified Korteweg-de Vries sine-Gordon equation versus other non-slowly-varying envelope-approximation models}, Physical Review A {\bf 79} (2009).
\bibitem{Leblond2011} H. Leblond and D. Mihalache, {\em Optical solitons in the fewcycle regime: recent theoretical results}, Romanian Reports in Physics {\bf 63} pp.1254-1266 (2011).
\bibitem{Ablowitz1981} M. J. Ablowitz and H. Segur, {\em Solitons and the inverse scattering transform}, SIAM  (1981).
\bibitem{Akhmediev2017} A. Ankiewicz and N. Akhmediev, {\em Rogue wave-type solutions of the mKdV equation and their relation to known NLSE rogue wave solutions}, Nonlinear Dyn {\bf 91} pp. 1931-1938 (2018).
\bibitem{Benjamin1967} B. T. Benjamin and J. E. Feir, {\em The disintegration of wave trains on deep water. Part 1. Theory}, J. Fluid Mech.  {\bf 27} pp. 417-430 (1967).
\bibitem{Ono1975} H. Ono, {\em Algebraic solitary waves in stratified fluids}, J. Phys. Soc. Japan {\bf 39} pp. 1082–1091 (1975).
\bibitem{Weideman1992} R. L. James and J. A. C. Weideman, {\em 
Pseudospectral Methods for the Benjamin-Ono Equation}, Advances in Computer Methods for Partial Differential Equations {\bf 7} (1992).
\bibitem{Boyd2011} J. P. Boyd and Z. Xu, {\em 
Comparison of three spectral methods for the Benjamin-Ono equation: Fourier psuedospectral, rational Christov functions and Gaussian radial basis functions}, Wave Motion {\bf 48} pp. 702-706 (2011).


\bibitem{Cole2025b} J. T. Cole and T. I. Johnson, {\em A split-step Christov method for approximating rational PDE solutions}, J. Comp.  Phys. (2025).






















































\end{thebibliography}

\end{document}